\documentclass[lettersize,journal]{IEEEtran}
\usepackage{amsmath,amsfonts}
\usepackage{algorithmic}
\usepackage{algorithm}
\usepackage{array}
\usepackage{amssymb}
\usepackage[caption=false,font=normalsize,labelfont=sf,textfont=sf]{subfig}
\usepackage{textcomp}
\usepackage{stfloats}
\usepackage{url}
\usepackage{verbatim}
\usepackage{graphicx}
\usepackage{makecell}
\usepackage{cite}
\usepackage{booktabs,multirow,xcolor}
\usepackage{threeparttablex}
\usepackage{tabularx}
\usepackage{wasysym} 
\definecolor{darkred}{RGB}{178,34,34}
\definecolor{darkgreen}{RGB}{0,100,0}

\newcommand{\cmark}{\checkmark}
\newcommand{\xmark}{\(\times\)}
\newcolumntype{C}[1]{>{\centering\arraybackslash}m{#1}} 
\newcolumntype{N}{>{\centering\arraybackslash}p{10mm}} %

\begin{document}

\title{MalRAG: A Retrieval-Augmented LLM Framework for Open-set Malicious Traffic Identification}

\author{Xiang Luo, Chang Liu, Gang Xiong, Chen Yang, Gaopeng Gou, Yaochen Ren, Zhen Li
\thanks{Xiang Luo, Chang Liu, Gang Xiong, Chen Yang, Gaopeng Gou, Yaochen Ren, Zhen Li are with the Institute of Information Engineering, Chinese Academy of Sciences, Beijing 10089, China, and also with the School of Cyber Security, University of Chinese Academy of Sciences, Beijing 100049, China. (email: \{luoxiang, liuchang, xionggang, yangchen, gougaopeng, renyaochen, lizhen\}@iie.ac.cn)}}

\markboth{Journal of \LaTeX\ Class Files,~Vol.~14, No.~8, August~2021}%
{Shell \MakeLowercase{\textit{et al.}}: A Sample Article Using IEEEtran.cls for IEEE Journals}


\maketitle

\begingroup
\renewcommand\thefootnote{}
\footnotetext{This work has been submitted to the IEEE for possible publication. Copyright may be transferred without notice, after which this version may no longer be accessible.}
\addtocounter{footnote}{0}
\endgroup

\begin{abstract}
Fine-grained identification of IDS-flagged suspicious traffic is crucial in cybersecurity. In practice, cyber threats evolve continuously, making the discovery of novel malicious traffic a critical necessity as well as the identification of known classes. Recent studies have advanced this goal with deep models, but they often rely on task-specific architectures that limit transferability and require per-dataset tuning.
\par In this paper we introduce MalRAG, the first LLM driven retrieval-augmented framework for open-set malicious traffic identification. MalRAG freezes the LLM and operates via comprehensive traffic knowledge construction, adaptive retrieval, and prompt engineering. Concretely, we construct a multi-view traffic database by mining prior malicious traffic from content, structural, and temporal perspectives. Furthermore, we introduce a Coverage-Enhanced Retrieval Algorithm that queries across these views to assemble the most probable candidates, thereby improving the inclusion of correct evidence. We then employ Traffic-Aware Adaptive Pruning to select a variable subset of these candidates based on traffic-aware similarity scores, suppressing incorrect matches and yielding reliable retrieved evidence. Moreover, we develop a suite of guidance prompts where task instruction, evidence referencing, and decision guidance are integrated with the retrieved evidence to improve LLM performance. Across diverse real-world datasets and settings, MalRAG delivers state-of-the-art results in both fine-grained identification of known classes and novel malicious traffic discovery. Ablation and deep-dive analyses further show that MalRAG effective leverages LLM capabilities yet achieves open-set malicious traffic identification without relying on a specific LLM.
\end{abstract}


\section{Introduction}
\begin{figure}
    \centering
    \includegraphics[width=\linewidth]{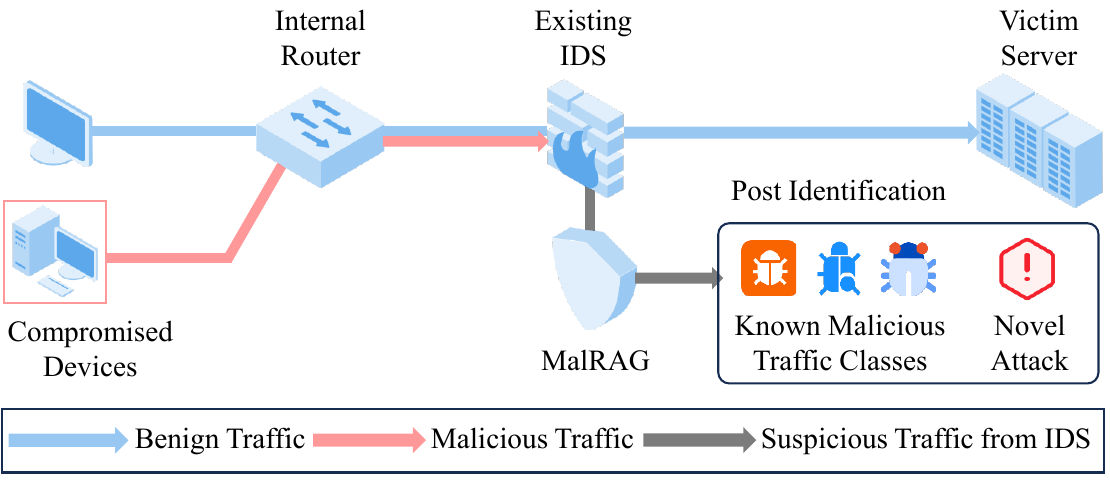}
    \caption{Illustration of the open-set malicious traffic identification task}
    \label{fig:topology}
\end{figure}

\IEEEPARstart{W}{ith} With the rapid evolution of network applications, malicious traffic has grown more dynamic and concealed, escalating operational risks and complicating network defense \cite{sommer2010outside, papadogiannaki2021survey}. Intrusion detection systems (IDSs) \cite{scarfone2007guide} provide first-line screening by issuing coarse alerts that extract a set of suspicious flows from massive network streams. However, these flagged flows require fine-grained identification to known classes\footnote{In the remainder of this paper, we refer to this fine-grained classification among known malicious classes simply as known malicious traffic identification for brevity.} for downstream response, as well as novelty discovery to uncover novel malicious traffic and inform defense adaptation. This combined requirement defines the open-set \textbf{M}alicious \textbf{T}raffic \textbf{I}dentification (MTI) problem \cite{luo2024identifying,meng2025detection}, as shown in Figure~\ref{fig:topology}. 

\par Existing work on open-set MTI has evolved along several lines. Early studies primarily addressed the fine-grained identification of known malicious traffic. Many relied on rule-based methods \cite{OISF_Suricata_UserGuide,roesch1999snort} or conventional machine learning \cite{taylor2016appscanner} with handcrafted features and protocol heuristics. These methods offered interpretability and efficiency, but they required substantial expert effort and scaled poorly in both accuracy and coverage. With the advent of deep learning, neural networks \cite{liu2019fs,deng2024net} and pre-trained models \cite{lin2022bert,zhou2025trafficformer} began to extract discriminative features directly from raw traffic, yielding notable gains in identification performance. Nevertheless, generalization to novel attacks remains a persistent limitation. To cope with newly emerging threats, more recent studies have explored schemes for novel malicious traffic discovery. In addition to learning patterns on known malicious traffic, they typically introduce additional detectors to capture potential novel patterns\cite{yang2021cade,luo2024identifying}. However, model architectures remain fragmented because they are tailored to specific tasks and datasets, thereby limiting transferability and necessitating per-task retraining. These constraints motivate a data-centric perspective that characterizes intrinsic traffic properties and reduces dependence on bespoke architectures and repeated training.

\par To exploit the shared characteristics of malicious traffic, we conduct an in-depth analysis of multiple datasets and distill two key observations. Firstly, flows within the same malicious traffic class exhibit consistent behavioral characteristics, which provides a stable basis for fine-grained identification. Secondly, novel attacks present discernible behavioral differences from established classes, which can assist in discovering such novel patterns in practice. Details are provided in Section \ref{data_analysis}. To this end, open-set MTI calls for models that generalize across tasks and can deeply mine discriminative characteristics of malicious traffic.

\par With advances in language modeling, large language models (LLMs) \cite{achiam2023gpt} combined with retrieval-augmented generation (RAG) \cite{lewis2020retrieval} provide a suitable foundation. LLMs offer large parameter capacity and a stable, reusable architecture that transfers across tasks, which reduces the need for per-task redesign\cite{brown2020language}. Meanwhile, RAG introduces external domain knowledge and task context via retrieval, enabling prompt-based adaptation to different domains without fine-tuning\cite{fan2024survey}. When it comes to network-traffic analysis, retrieval is able to assemble traffic characteristics from complementary views, and the LLM can use these characteristics to assign suspicious flows to specific malicious categories and to flag novel patterns that diverge from established behavior. This combination reduces dependence on bespoke architectures and repeated retraining while aligning the approach with the discriminative properties of the data.

\par Therefore, we present MalRAG, the first LLM-driven retrieval-augmented framework for open-set malicious traffic identification that operates without fine-tuning. MalRAG consists of a multi-view traffic database and 3 modules: Prompt Construction, Adaptive Retrieval, and Answer Generation. The traffic database stores flow information from different classes of known malicious traffic across complementary views, capturing content, structural and temporal characteristics to support retrieval. Once the database is in place, we introduce an Adaptive Retrieval module to supply the LLM with flow information relevant to the query. Specifically, given a suspicious flow, we first conduct Coverage-Enhanced Retrieval to improve retrieval completeness and obtain an initial top-$k$ set of evidence. We then apply Traffic-Aware Adaptive Pruning using similarity thresholds to further reduce false evidence induction. In parallel, we design a suite of guidance prompts in the Prompt Construction module that covers task instruction, evidence referencing, and decision guidance. The guidance prompt and retrieved evidence collectively enhances LLM’s understanding of the problem and helps it deliver reliable answers in the Answer Generation module. 
\par Our main contributions are summarized as follows:
\begin{itemize}
    \item We introduce MalRAG, the first LLM-driven, retrieval-augmented framework for open-set malicious traffic identification. By grounding a frozen LLM in a multi-view traffic database with comprehensive knowledge, MalRAG supports both accurate identification of known classes and discovery of novel malicious traffic.
    \item To improve retrieval completeness, we propose Coverage-Enhanced Retrieval which queries multiple views to assemble the most relevant malicious traffic samples as initial evidence. Moreover, Traffic-Aware Adaptive Pruning is adopted to adaptively refine the evidence with similarity checks, thereby increasing evidence precision for downstream identification.
    \item We design a suite of guidance prompts which comprises task instruction, evidence referencing, and decision guidance, to improve the LLM’s contextual understanding of retrieved evidence and the quality of its responses for open-set malicious traffic identification.
    \item We evaluate MalRAG on diverse real-world datasets, and show that MalRAG outperforms state-of-the-art methods on both known classes identification and novel malicious traffic discovery. Ablations and deep-dive analyses indicate that MalRAG leverages LLM capabilities while remaining largely backbone-agnostic, rather than relying on any specific LLM.
\end{itemize}

\section{Related Work}
In this section, we review prior studies related to open-set malicious traffic identification. Existing research can be broadly categorized into three lines: (1) methods for identifying known malicious traffic, (2) methods for discovering novel malicious traffic, and (3) recent attempts to leverage LLMs for traffic analysis. A summary of representative works is provided in Table \ref{tab:related_work}.

\subsection{Known Malicious Traffic Identification}
MalRAG targets multi-class malicious traffic identification, going beyond binary benign–malicious detection. Although many existing works \cite{liu2021jaqen,xing2021ripple,DBLP:conf/icacci/VinayakumarSP17c,DBLP:journals/bigdata/Al-TuraikiA21,alrayes2024cnn,sun2024gnn,pujol2022unveiling} focus on binary classification of flows, in this section we primarily review approaches for fine-grained identification among different malicious behaviors or families. Early efforts in this space are rule- and signature-based systems such as Snort~\cite{roesch1999snort} and Suricata~\cite{OISF_Suricata_UserGuide}, which perform detailed packet inspection using predefined rules. While these systems offer high interpretability and low computational cost, they depend heavily on expert-defined signatures and struggle to detect encrypted or evolving traffic patterns.
\par To overcome the rigidity of manual rules, researchers turned to machine learning (ML) based approaches. Early works such as AppScanner\cite{taylor2016appscanner} and FlowLens\cite{barradas2021flowlens} relied on handcrafted statistical and protocol features combined with traditional classifiers like random forests to identify malicious behaviors. While these techniques improved accuracy over rule systems, they required substantial expert effort and were sensitive to feature design. Later studies adopted neural architectures to extract higher-level representations directly from traffic. For instance, DF\cite{sirinam2018deep} and HAST-IDS\cite{wang2017hast} used convolutional models to capture spatial correlations, while FS-Net\cite{liu2019fs} and PARALLEL-LSTM\cite{niu2022uncovering} modeled sequential dependencies within flow dynamics. In addition, CBSeq\cite{cui2023cbseq} and AN-Net\cite{deng2024net} incorporated attention mechanisms to highlight salient temporal patterns; and ST-Graph\cite{fu2022encrypted} and MalDiscovery\cite{hong2023graph} represented relational and topological dependencies among flows with graph structures. Despite their improved accuracy, their limited parameter capacity hinders their generalization across diverse scenarios.
\par More recently, pretrained models have been introduced to leverage vast amounts of unlabeled traffic data. ET-BERT \cite{lin2022bert} and TrafficFormer \cite{zhou2025trafficformer} adopted self-supervised pretraining on packet or flow sequences to learn general representations that can be fine-tuned for downstream classification tasks. Combining token and packet level attention with PRPP and FCL pretraining, MIETT\cite{chen2025miett} learns flow-aware representations for encrypted traffic and achieves competitive performance. On the other hand, some traffic pretraining works adopt a GPT-style paradigm: NetGPT\cite{meng2023netgpt} textualizes flows for unified understanding-and-generation with prompt adaptation, while TrafficGPT\cite{qu2024trafficgpt} employs linear attention for long-flow classification and realistic synthesis. Morever, YaTC \cite{zhao2023yet} and Flow-MAE \cite{hang2023flow} applied masked autoencoding to model traffic semantics, while TFE-GNN \cite{zhang2023tfe} modeled raw packet bytes as graphs and used a GNN-based encoder with dual embedding and fusion to learn encrypted traffic representations. These methods substantially improve performance with limited labeled data. However, a recent systematization of knowledge study \cite{wickramasinghe2025sok} revealed that many pretrained traffic classifiers tend to rely on shortcut patterns or strong payload-related features, which may not generalize across datasets or encryption protocols. As a result, such approaches still require task-specific fine-tuning and suffer from the closed-world assumption, limiting their adaptability to novel malicious traffic.

\begin{table}[t]
\centering
\caption{Comparison of representative methods for open-set malicious traffic identification.}
\label{tab:related_work}
\begin{threeparttable}
\renewcommand{\arraystretch}{1.35}
\setlength{\tabcolsep}{2pt}
\scriptsize
\begin{tabularx}{\linewidth}{@{}l|X|c|c|l|c@{}}
\toprule
\textbf{Type} & \textbf{Method} & \textbf{KFGI} & \textbf{NMTD} & \textbf{Arch} & \textbf{Train}\\
\midrule
\multirow{2}{*}{\makecell{\textbf{Rule-} \\ \textbf{based}}}%
& Jaqen\cite{liu2021jaqen}, Ripple\cite{xing2021ripple} & \xmark & \xmark & Rules & \Circle \\
& Snort\cite{roesch1999snort}, Suricata\cite{OISF_Suricata_UserGuide} & \cmark & \xmark & Rules & \Circle \\
\midrule

\multirow{10}{*}{\makecell{\textbf{ML-} \\ \textbf{based}}}%
& APPScanner\cite{taylor2016appscanner}, FlowLens\cite{barradas2021flowlens} & \cmark & \xmark & Forest & \CIRCLE \\
& CNN-based IDS\cite{DBLP:conf/icacci/VinayakumarSP17c,DBLP:journals/bigdata/Al-TuraikiA21,alrayes2024cnn} & \xmark & \xmark & CNN & \CIRCLE \\
& DF\cite{sirinam2018deep}, HAST\cite{wang2017hast} & \cmark & \xmark & CNN & \CIRCLE \\
& FS\textendash Net\cite{liu2019fs}, \cite{niu2022uncovering} & \cmark & \xmark & RNN & \CIRCLE \\
& CBSeq\cite{cui2023cbseq}, AN\textendash Net\cite{deng2024net} & \cmark & \xmark & Transformer & \CIRCLE \\
& GNN-IDS\cite{sun2024gnn,pujol2022unveiling} & \xmark & \xmark & Graph & \CIRCLE
\\
& ST\textendash Graph\cite{fu2022encrypted}, MalDiscovery\cite{hong2023graph} & \cmark & \xmark & Graph & \CIRCLE \\
& ET-BERT\cite{lin2022bert}, MIETT\cite{chen2025miett}, \cite{zhou2025trafficformer} & \cmark & \xmark & Transformer & \LEFTcircle \\
& YaTC\cite{zhao2023yet}, Flow-MAE\cite{hang2023flow} & \cmark & \xmark & MAE & \LEFTcircle\\
& TrafficGPT\cite{qu2024trafficgpt}, NetGPT\cite{meng2023netgpt} & \cmark & \xmark & GPT & \LEFTcircle \\
& CADE\cite{yang2021cade}, ICE\textendash CP\cite{luo2024identifying} & \cmark & \cmark & DNN & \CIRCLE \\
& ZTI\cite{jin2020zero}, GradDB\cite{yang2021deep}, \cite{xia2021gmaf,han2024ecnet} & \cmark & \cmark & CNN & \CIRCLE \\
\midrule

\multirow{2}{*}{\makecell{\textbf{LLM-} \\ \textbf{based}}}%
& TrafficLLM\cite{cui2025trafficllm}, MOTA\cite{li2025mota} & \cmark & \xmark & LLM & \LEFTcircle \\
& \textbf{MalRAG (Ours)} & \cmark & \cmark & LLM & \Circle \\
\bottomrule
\end{tabularx}

\begin{tablenotes}[flushleft]\footnotesize
\item[] \textit{Notes.} KFGI = known fine\textendash grained identification; NMTD = novel malicious traffic discovery; Arch = Architecture; 
Train marks: \CIRCLE\ (train), \LEFTcircle\ (fine-tune), \Circle\ (neither).
“ML” denotes \textbf{Machine Learning}. For lengthy method names, only citations are kept.
\end{tablenotes}
\end{threeparttable}
\end{table}

\subsection{Novel Malicious Traffic Discovery}
\par To enable the detection of previously unseen threats, recent studies have extended malicious traffic classifiers with mechanisms for identifying samples that deviate from known distributions. According to the metric used for novelty assessment, existing methods can be broadly divided into three categories: reconstruction-based, gradient-based, and distance-based approaches.
\par Reconstruction-based methods assume that models trained on known data fail to reconstruct unseen traffic patterns. For instance, ZTI\cite{jin2020zero} detects novel malicious traffic using an auxiliary autoencoder trained on known flows, and samples with high reconstruction errors are flagged as potential novel malicious traffic. On the other hand, gradient-based methods assess the model’s sensitivity to new inputs through back-propagation gradients. Specifically, GradDB\cite{yang2021deep} identifies novel malicious traffic by examining the magnitude of gradients during inference, as known samples produce stable, low-magnitude gradients, whereas novel samples yield irregular and larger ones. GMAF\cite{xia2021gmaf} further enhances this paradigm by incorporating an additive angular margin (ArcFace) loss during training.
\par Different from the above-mentioned methods, Distance-based methods determine novelty by measuring the distance between a test sample and class prototypes in the learned feature space. CADE\cite{yang2021cade} and ICE-CP\cite{luo2024identifying} compute distances from the test sample to each known malicious class center, and if all distances exceed a threshold, the sample is determined as potential novel malicious traffic. However, these methods typically rely on manually tuned thresholds and are sensitive to intra-class distribution variations.
\par Although these approaches demonstrate the feasibility of discovering novel malicious traffic, their reliance on static thresholds and single-view representations still limits robustness and adaptability in dynamic network environments.

\subsection{LLMs for Traffic Analysis}
Although LLM-based have been widely explored in other security domains \cite{dinis2025large,DBLP:conf/ndss/LinM25}, their use in traffic analysis is still in its early stages. Existing attempts are limited to a few preliminary frameworks, most notably TrafficLLM\cite{cui2025trafficllm} and MoTA\cite{li2025mota}. TrafficLLM introduces traffic-domain tokenization and a dual-stage fine-tuning pipeline to learn generic representations from heterogeneous raw traffic, enabling both detection and generation with strong generalization to unseen flows. MoTA adopts a lightweight mixture-of-agents architecture to fine-tune mainstream LLMs for diverse classification scenarios, leveraging agent collaboration to enhance robustness under mixed noise.
\par However, both approaches require task-specific fine-tuning, incurring nontrivial computational and maintenance costs. In addition, they also rely heavily on payload content, and prior studies\cite{wickramasinghe2025sok} show that strong payload cues can act as shortcuts that hinder cross-domain generalization.

\section{Problem Statement}
In this work, we assume the availability of a prior corpus of labeled malicious traffic covering a set of known malicious classes. As shown in Figure \ref{fig:topology}, this corpus consists of raw traffic data, from which we can extract multiple complementary feature views (e.g., payload content, packet-length sequences, and inter-arrival times) to construct a traffic knowledge base. In deployments where only pre-computed features are stored instead of raw traffic, our analysis is naturally restricted to matching suspicious flows against the available feature views. 

\par Under this premise, we formulate the task as an open-set malicious traffic identification (MTI) problem: for each flow flagged as suspicious by the IDS, we aim to determine whether it (1) corresponds to malicious traffic from one of the known classes represented in the prior corpus, or (2) constitutes novel malicious traffic that is not covered by this corpus.

\par Formally, let $\mathcal{X} = \{x_i\}_{i=1}^N$ denote the set of suspicious flows returned by the IDS. 
Each flow $x_i$ may include one or more feature views, such as packet-length sequences, payload bytes, or inter-arrival time intervals, depending on data availability.
We define the known malicious traffic classes as $\mathcal{C}_{\text{mal}} = \{c_1, c_2, \dots, c_M\}$. 
In addition, $\mathcal{C}_{\text{nov}}$ represents novel malicious traffic classes that are not represented in the prior corpus and have not been observed during training.

The goal of open-set MTI is to learn a mapping function
$
f: \mathcal{X} \rightarrow \mathcal{C}_{\text{mal}} \cup \mathcal{C}_{\text{nov}},
$
where each suspicious flow $x_i \in \mathcal{X}$ is assigned either to a known malicious class or to a potential novel malicious traffic.

\section{Key Observation}
\label{data_analysis}
\begin{figure}[t]
\centering
    \subfloat[Packet Length]{
    \includegraphics[width=0.45\linewidth]{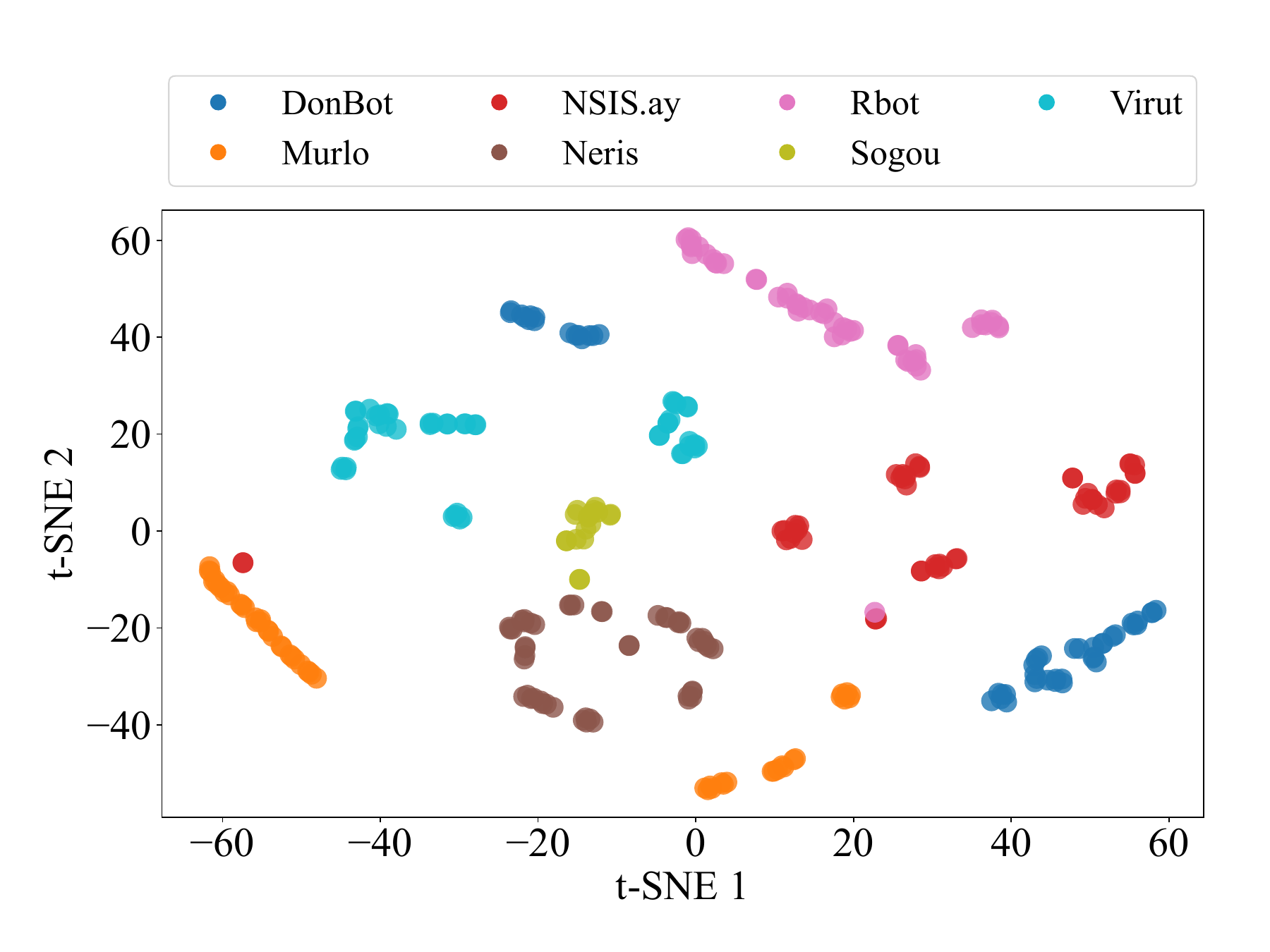}
    \label{fig:analyze_pkt}
    }
    \subfloat[Payload]{\includegraphics[width=0.45\linewidth]{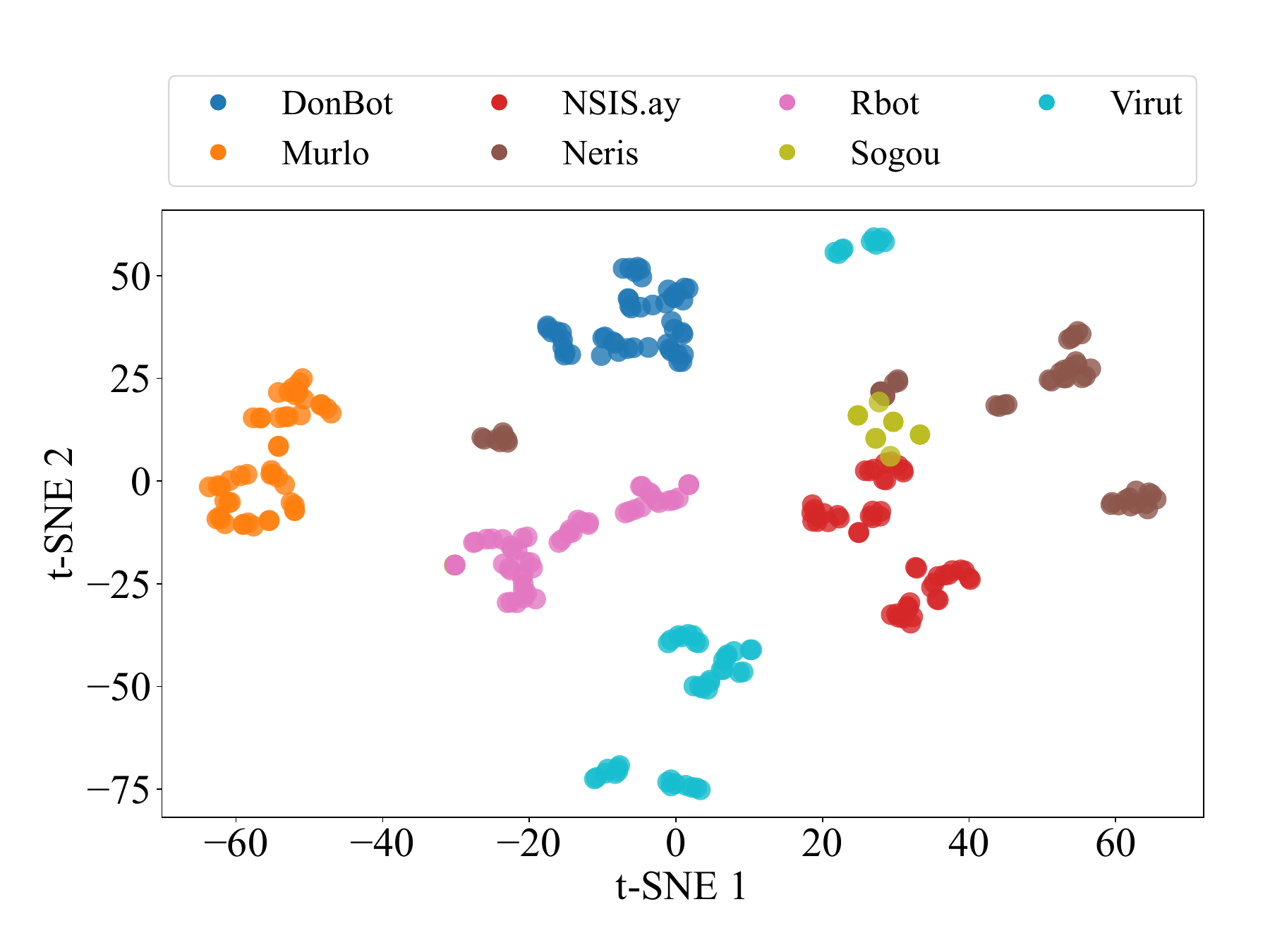}
    \label{fig:analyze_payload}
    }
    \caption{t-SNE visualization of packet length and payload features across malicious traffic classes in CTU-13 dataset}
    \label{fig:analyze_01}
\end{figure}

\begin{figure}
\centering
 \subfloat[Packet length]{
  \includegraphics[width=0.45\linewidth]{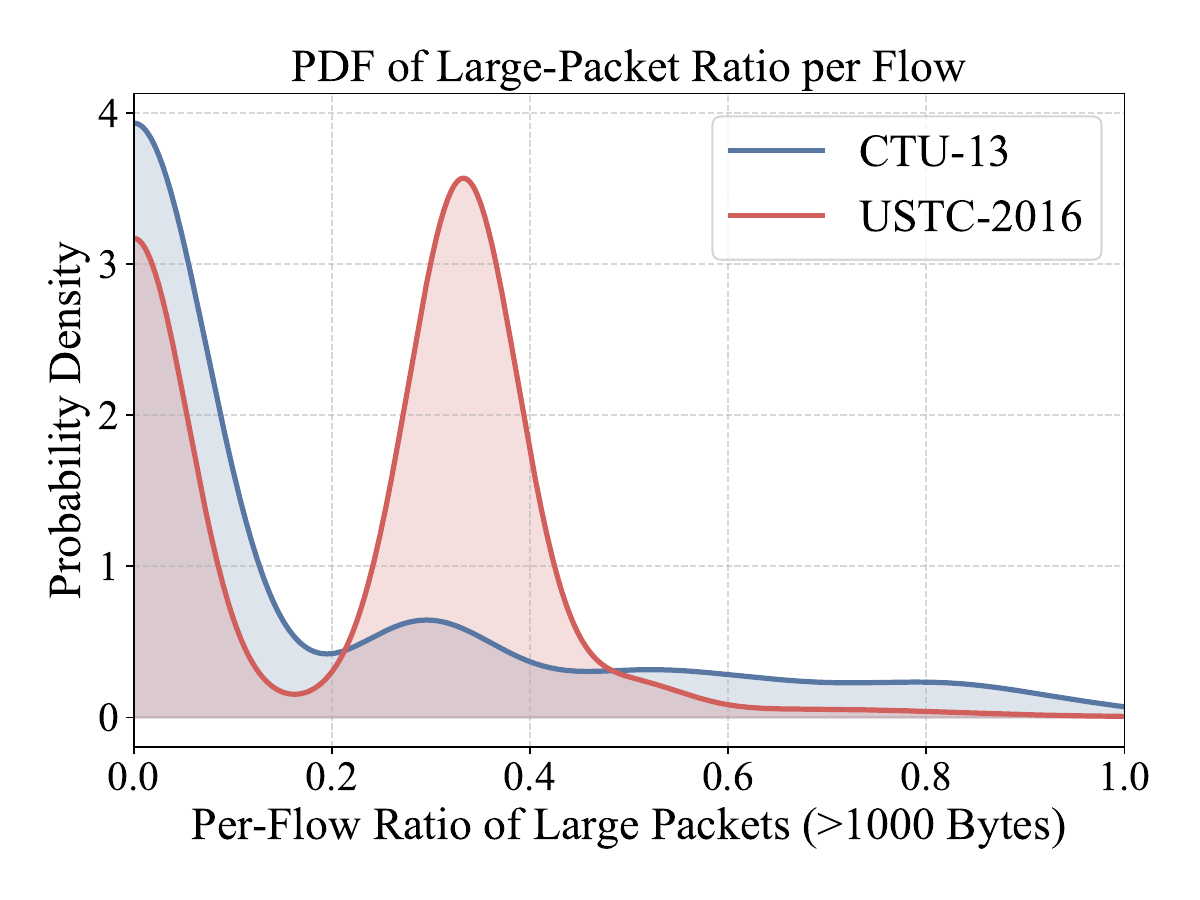}
  \label{fig:analyze_02_cdf_packet.pdf}
 }
 \subfloat[Arrival time]{
  \includegraphics[width=0.45\linewidth]{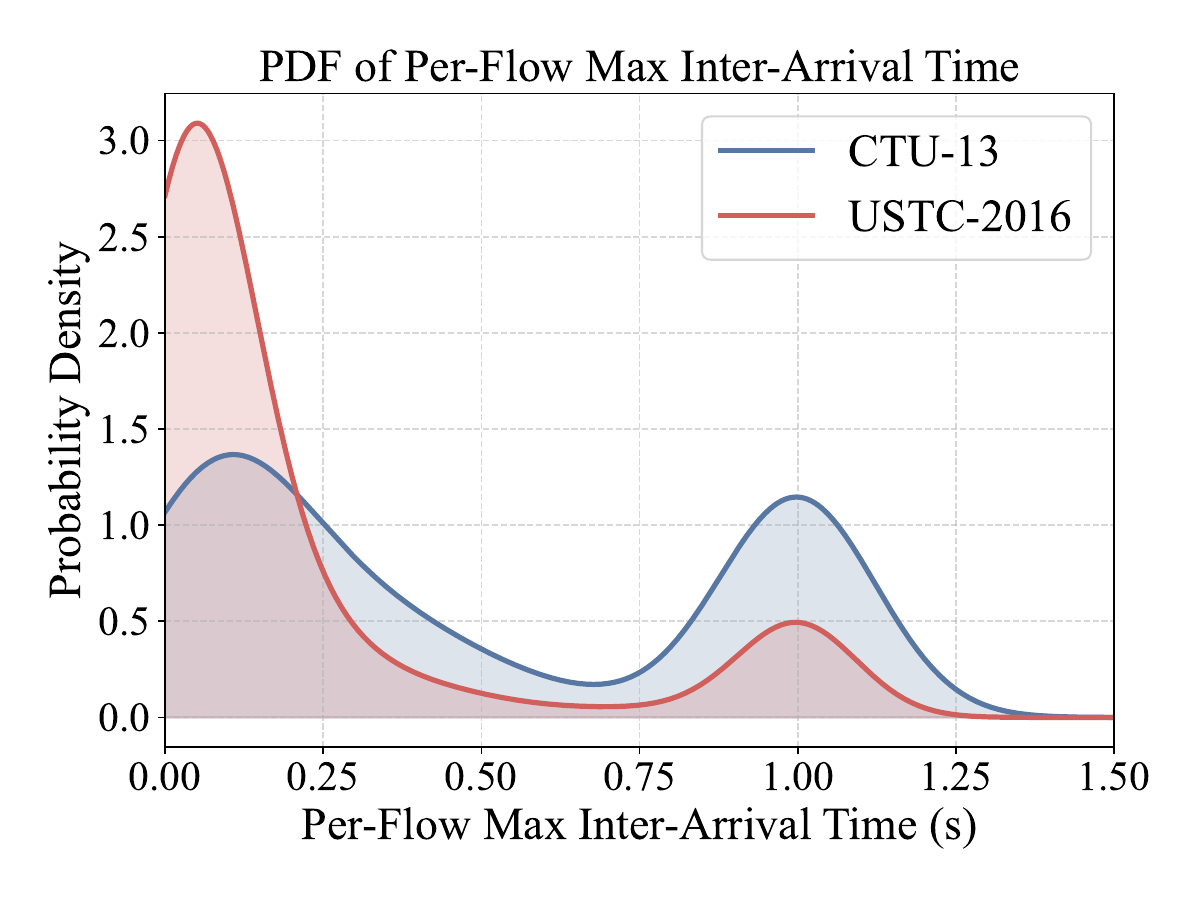}
  \label{fig:analyze_02_cdf_time.pdf}
 }
 \caption{PDFs of packet length and inter-arrival time across malicious datasets: CTU-13 (known) vs. USTC-2016 (novel)}
 \label{fig:analyze_02}
\end{figure}

To better understand intrinsic malicious traffic characteristics and support open-set MTI, we analyze the characteristics of payloads, packet-length sequences, and inter-arrival time intervals, and summarize two observations across datasets and traffic classes, as shown in Figure\ref{fig:analyze_01} and Figure\ref{fig:analyze_02}. 
\par\textbf{(1) Intra-class Consistency in the same class of malicious traffic}. Malicious traffic belonging to the same family exhibits strong intra-class consistency. To examine this phenomenon, we analyze packet-length sequences and inter-arrival times as structural indicators of encrypted traffic. As shown in Figure \ref{fig:analyze_01}, the t-SNE visualization of packet-length sequences and payload data from the CTU-13 dataset\cite{CTU-13} reveals that samples from the same malicious family cluster tightly with coherent local geometry, while distinct families remain clearly separated. These class-informative patterns demonstrate that the packet-length and payload perspectives provide a stable and discriminative basis for relevant traffic retrieval.

\par\textbf{(2) Novel malicious traffic exhibits noticeable distributional divergence from prior known malicious traffic.} we compare the probability density functions (PDFs) of packet lengths and inter-arrival times between CTU-13 and USTC-2016~\cite{wang2017malware} datasets, using them as representative corpora of known and novel malicious traffic, respectively. As shown in Figure \ref{fig:analyze_02}, the distributions of USTC-2016 systematically deviate from those of CTU-13 in both feature views, revealing clear shifts driven by the emergence of newer attack behaviors. These shifts highlight the importance of leveraging complementary feature views to obtain a more complete characterization of contemporary malicious patterns and to enhance robustness against temporal drift.


\begin{figure*}
    \centering
    \includegraphics[width=\textwidth]{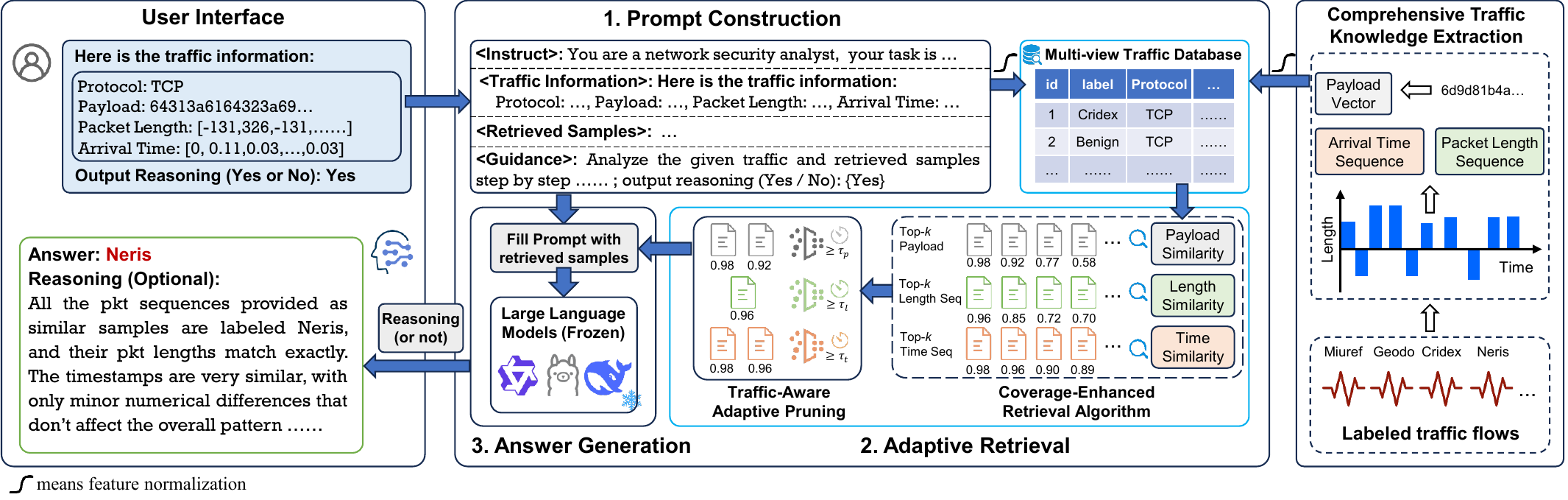}
    \caption{Overall workflow of MalRAG}
    \label{fig:framework}
\end{figure*}

\section{Methodology}
To achieve the goal of effective open-set MTI under a unified model architecture without frequent tuning, we propose MalRAG, the first LLM-driven retrieval-augmented framework. It builds upon a pre-constructed multi-view traffic database that stores representative flow metadata collected from labeled malicious traffic data. Our method targets 2 operational objectives in a unified pipeline: (i) accurate identification of \emph{known} malicious traffic, (ii) effective discovery of \emph{novel} malicious traffic. 

\subsection{Framework Overview}
MalRAG’s overall framework (Figure~\ref{fig:framework}) combines a multi-view traffic database with three modules: Prompt Construction, Adaptive Retrieval, and Answer Generation.

\par Given a suspicious flow flagged by an IDS, MalRAG first constructs a structured guidance prompt through the \textbf{Prompt Construction} module. Following a unified template, this prompt encodes the suspicious flow as normalized traffic information and overlays three types of guidance: task instruction that specifies the objective, evidence referencing that anchors the retrieved samples and labels, and decision guidance that helps the LLM better use this evidence to decide between known classes and novel malicious traffic.

\par Next, the \textbf{Adaptive Retrieval} module performs a two-stage process to retrieve supporting evidence from the database. In the first stage, MalRAG conducts \emph{Coverage-Enhanced Retrieval}, searching across different feature dimensions to obtain the top-$k$ most relevant samples and their labels as evidence. In the second stage, MalRAG performs \emph{Traffic-Aware Adaptive Pruning} to filter out evidence with low similarity to the query, ensuring that only reliable evidence is finally inserted into the final prompt.
\par Finally, the \textbf{Answer Generation} module feeds the completed prompt into the LLM to generate the final output. 
Depending on the user’s reasoning option, MalRAG can either produce the predicted result alone or provide an evidence-grounded reasoning explanation. 
This retrieval-augmented workflow enables MalRAG to achieve interpretable, adaptable, and fine-tuning-free malicious traffic identification.

\subsection{Comprehensive Traffic Knowledge Extraction}
\label{sec:database_construction}
Current open-source LLMs are not designed for network traffic analysis and thus lack domain-specific knowledge of malicious behaviors or protocol-level patterns. To bridge this gap, we construct a multi-view traffic database that serves as the external comprehensive knowledge base for MalRAG. This database provides the model with structured, retrievable representations of diverse traffic behaviors, enabling contextual reasoning grounded in real-world examples rather than general linguistic priors.
\par\textbf{Data Composition.}
From the perspectives of traffic content, structural patterns, and temporal behavior, we extract three primary types of raw traffic information from flow metadata: payload bytes, packet-length sequences, and inter-arrival time sequences, together with the corresponding class label. These complementary views jointly capture key aspects of malicious behavior and provide a comprehensive basis for subsequent similarity retrieval and analysis.

\par\textbf{Feature Normalization.} 
To ensure cross-sample comparability and robustness, all traffic samples are converted into standardized vector forms. For payload data, we first remove strong identifiers using the strong-feature randomization strategy in \cite{wickramasinghe2025sok}, preventing the model from overfitting to brittle payload artifacts. The remaining hexadecimal bytes are then truncated to a fixed length $L_{\text{pay}}$ and converted to an integer vector with length normalization.
\par To better capture periodic, bursty, and patterns in structural and temporal series of traffic, we further transform the packet-length and inter-arrival time sequences into a form amenable to frequency-domain analysis. Specifically, the raw packet-length and inter-arrival time sequences are first normalized to fixed lengths $L_{\text{len}}$ and $L_{\text{time}}$, respectively, using a truncation operator that either shortens or zero-pads the sequences. For either normalized sequence $v \in \{v^{(\text{len})}, v^{(\text{time})}\}$ with length $L$, we then segment it into non-overlapping frames of window size $W_{\text{seg}}$:
\begin{equation}
\small
    \begin{aligned}
        f_i &= v[((i-1) \times W_{\text{seg}}) : (i \times W_{\text{seg}})], \quad 1 \le i \le N_f, \\
        N_f &= \Bigg\lceil \frac{L}{W_{\text{seg}}} \Bigg\rceil,
    \end{aligned}
\end{equation}
where $N_f$ denotes the number of frames. For each frame $f_i$, we perform a Discrete Fourier Transformation (DFT) 
to capture its frequency-domain characteristics, following prior work~\cite{fu2021realtime}:
\begin{equation}
\small
    F_{ik} = \sum_{n=1}^{W_{\text{seg}}} f_i[n] 
\, e^{-j 2\pi (n-1)(k-1)/W_{\text{seg}}}, 
\quad 1 \le k \le W_{\text{seg}}.
\end{equation}
Each complex coefficient $F_{ik} = a_{ik} + j b_{ik}$ is converted into a real-valued amplitude:
\begin{equation}
\small
    p_{ik} = \sqrt{a_{ik}^2 + b_{ik}^2}.
\end{equation}
We retain the first half of the spectrum to eliminate conjugate redundancy, defining
\begin{equation}
\small
    P_i = [p_{i1}, \dots, p_{iK_f}], \quad K_f = \Big\lfloor \frac{W_{\text{seg}}}{2} \Big\rfloor.
\end{equation}
Finally, we aggregate the per-frame spectral vectors via mean pooling to obtain a 
global frequency-domain representation:
\begin{equation}
\small
   \bar{P} = \frac{1}{N_f} \sum_{i=1}^{N_f} P_i \in \mathbb{R}^{K_f}. 
\end{equation}

We apply this process to both packet-length and inter-arrival time sequences, obtaining $\bar{P}^{(\text{len})}$ and $\bar{P}^{(\text{time})}$. The time-domain vectors ($v^{(\text{len})}, v^{(\text{time})}$) and frequency-domain vectors ($\bar{P}^{(\text{len})}, \bar{P}^{(\text{time})}$) are all stored in the database for later retrieval.

\par This database serves as the foundational knowledge corpus for MalRAG. During inference, retrieved samples from this corpus provide traffic-aware evidence that supplements the LLM’s general reasoning capability. 
Because the LLM remains frozen, the knowledge base can be easily updated or expanded to incorporate new attack patterns without retraining, ensuring adaptability to evolving network environments.

\subsection{Prompt Construction}
\begin{figure}
    \centering
    \includegraphics[width=0.48\textwidth]{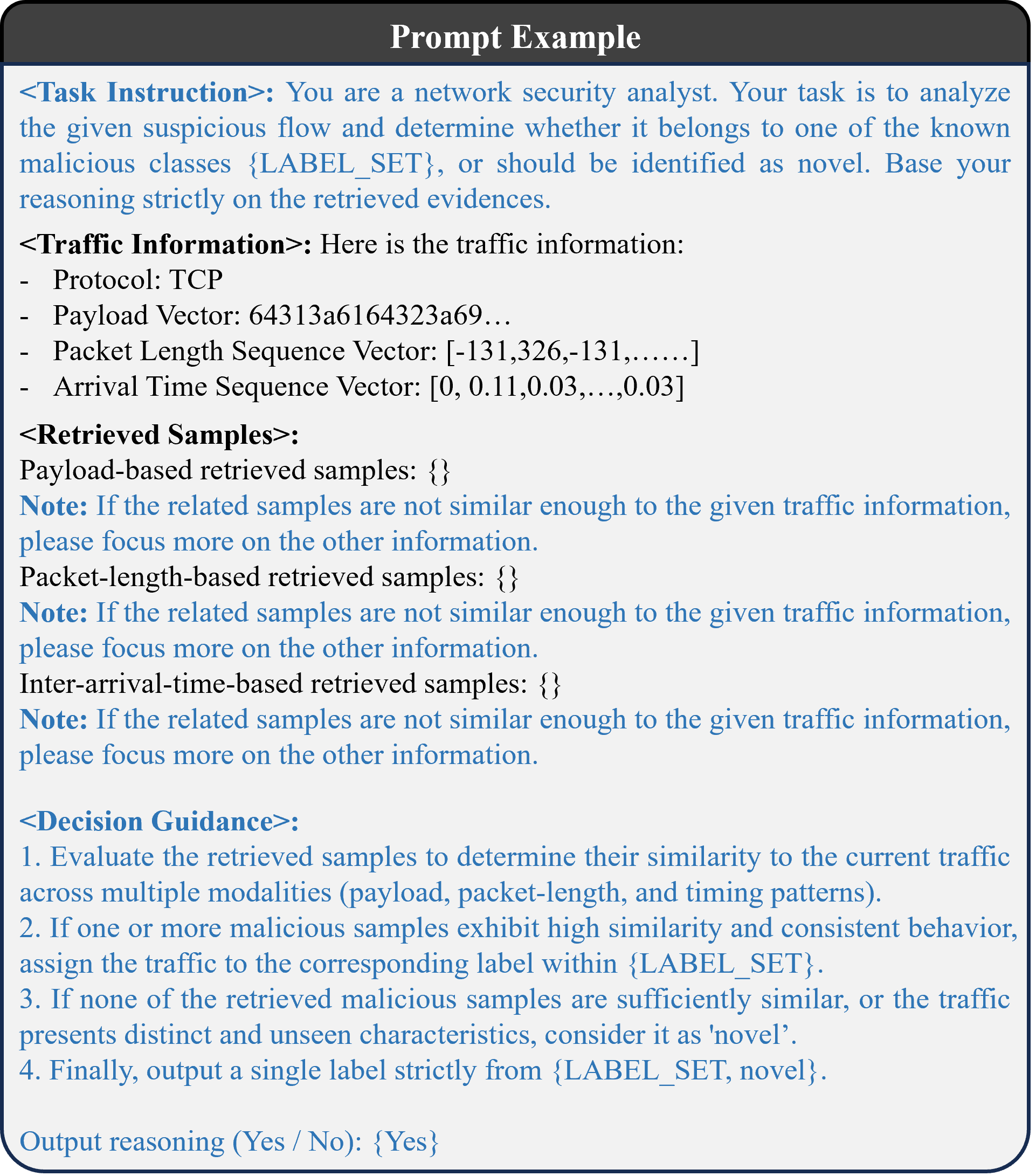}
    \caption{Details of Prompt Construction. The \textcolor{blue}{blue} sections in the figure represent the prompt designed by us.}
    \label{fig:prompt_example}
\end{figure}

Prompt construction bridges the given traffic data and the large language model. To help the LLM better interpret the provided traffic information, we design the guidance prompt around three key aspects: task understanding, context comprehension, and answer generation. The prompt is organized into four structured segments, as illustrated in Fig.~\ref{fig:prompt_example}: \textit{Task Instruction}, \textit{Traffic Information}, \textit{Retrieved Samples}, and \textit{Decision Guidance}.

\par\textbf{Task Instruction.} This segment is part of the prompt we design, explicitly defining the LLM’s analytical role and task objective. The large language model is instructed to act as a network security analyst, tasked with analyzing a given suspicious flow and determining whether it belongs to a known malicious class, or represents a novel attack. To ensure consistent and bounded outputs, we restrict the answer space $\mathcal{C}$ to $\mathcal{C} = \text{LABEL\_SET} \cup \left\{ \text{novel} \right\}$, where \text{LABEL\_SET} denotes the set of all known traffic categories stored in the constructed database. These categories correspond to the labeled classes of malicious and benign traffic available during database construction, defining the scope of known knowledge for subsequent reasoning and guiding the model towards accurate, task-specific decisions.

\par\textbf{Traffic Information.} This segment embeds the traffic data provided by the user, with the included views determined by data availability. The input may contain normalized representations of the payload bytes, the packet-length sequence, and the inter-arrival-time sequence for the given flow. When a particular view is unavailable, its field in the prompt is left empty. All included features are formatted as numerical arrays to ensure compatibility with textual input.

\par\textbf{Retrieved Samples.}  
This segment is initially constructed as a placeholder during the prompt formulation stage, as the actual retrieval has not yet been performed. Each placeholder corresponds to evidence from a different feature view, including payload, packet-length, and inter-arrival time. To assist the LLM in understanding the context, we augment each placeholder with a evidence referencing **Note** that provides instructions on how to interpret the retrieved evidence. After the adaptive retrieval process, these placeholders are dynamically populated with the retrieved samples and their associated labels from the traffic database. This design ensures that each feature view contributes relevant contextual evidence, helping LLM better understand the traffic data and making the analysis more coherent.

\par\textbf{Decision Guidance.} This segment is a key component of the prompt we design, providing explicit instructions to guide the LLM’s reasoning and final decision. It first directs the LLM to analyze whether the retrieved samples are sufficiently similar to the given traffic data and assess if their labels align with other flow characteristics such as payload, protocol, and timing patterns. If none of the retrieved samples appear relevant or consistent with the given traffic information, the model is instructed to classify the input as \emph{novel}, signaling the need for further analysis. Finally, the guidance specifies that the model must select one label from the set \(\{\text{LABEL\_SET}, \text{novel}\), and generate reasoning output only if the user’s reasoning option is enabled. This structured guidance enhances the LLM’s ability to make informed and context-aware decisions.

\subsection{Adaptive Retrieval}
Adaptive Retrieval takes the user-provided suspicious flow as input, and outputs a set of labeled candidates flows that will serve as contextual evidence for the LLM. To make this evidence both accurate and relevant, we employ two key techniques. First, Coverage-Enhanced Retrieval improves precision by evaluating correlations across multiple feature views, rather than relying on a single space. Second, Traffic-Aware Adaptive Pruning filters out irrelevant or erroneous evidence using view-specific similarity thresholds and cross-view checks, retaining only the most reliable samples. The overall procedure is summarized in Algorithm~\ref{alg:adaptive_retrieval}, with detailed components described next.
\begin{algorithm}[t]
\scriptsize
\caption{Adaptive Retrieval}
\label{alg:adaptive_retrieval}
\begin{algorithmic}[1]
\REQUIRE Query flow $x_q$; modality set $\mathcal{M} = \{\text{payload}, \text{length}, \text{time}\}$;
        traffic DB $\mathcal{D}$ with per-view vectors $x_i^{(m)}$, protocol $p_i$, class $c_i$;
        distances $d_H$ (payload) and $d_E$ (length/time); top-$k$; tolerance $\alpha$.
\ENSURE Refined per-view evidence $\{\mathcal{R}^{(m)}_{\mathrm{ref}}\}_{m\in\mathcal{M}}$
        and unified pool $\mathcal{E}$.

\STATE \COMMENT{Part A: Coverage-Enhanced Retrieval (initial screening)}
\FOR{each modality $m \in \mathcal{M}$}
    \STATE Obtain query representation $x_q^{(m)}$ from $x_q$.
    \STATE Determine protocol-constrained candidate set $\mathcal{D}^{(m)}_{\mathrm{proto}}(x_q)$.
    \IF{$m = \text{payload}$}
        \STATE $d \leftarrow d_H$.
    \ELSE
        \STATE $d \leftarrow d_E$.
    \ENDIF
    \STATE Compute $d\!\left(x_q^{(m)}, x_i^{(m)}\right)$ for all $x_i^{(m)} \in \mathcal{D}^{(m)}_{\mathrm{proto}}$.
    \STATE $\widehat{\mathcal{R}}^{(m)} \leftarrow \text{Top-}k_{\min}\{\,d\!\left(x_q^{(m)}, x_i^{(m)}\right)\,\}$ with labels $(c_i, p_i)$.
\ENDFOR

\STATE
\STATE \COMMENT{Part B: Traffic-Aware Adaptive Pruning (filtering)}
\STATE \COMMENT{Class–protocol statistics $(\bar d^{(m)}_{c,p}, \sigma^{(m)}_{c,p})$ can be precomputed and cached.}

\FOR{each modality $m \in \mathcal{M}$}
    \STATE $\mathcal{R}^{(m)}_{\mathrm{ref}} \leftarrow \emptyset$.
    \FOR{each $x_i^{(m)} \in \widehat{\mathcal{R}}^{(m)}$ with $(c_i, p_i)$}
        \STATE $\tau^{(m)}_{c_i,p_i} \leftarrow \bar d^{(m)}_{c_i,p_i} + \alpha \cdot \sigma^{(m)}_{c_i,p_i}$.
        \IF{$d\!\left(x_q^{(m)}, x_i^{(m)}\right) \le \tau^{(m)}_{c_i,p_i}$}
            \STATE $\mathcal{R}^{(m)}_{\mathrm{ref}} \leftarrow \mathcal{R}^{(m)}_{\mathrm{ref}} \cup \{x_i^{(m)}\}$.
        \ENDIF
    \ENDFOR
\ENDFOR

\STATE $\mathcal{E} \leftarrow \bigcup_{m\in\mathcal{M}} \mathcal{R}^{(m)}_{\mathrm{ref}}$.
\RETURN $\{\mathcal{R}^{(m)}_{\mathrm{ref}}\}_{m\in\mathcal{M}}, \mathcal{E}$.
\end{algorithmic}
\end{algorithm}

\subsubsection{Coverage-Enhanced Retrieval}
To improve the coverage of relevant evidence, MalRAG performs Coverage-Enhanced Retrieval, which searches across multiple feature views to identify similar samples as evidence. Given the user-provided traffic information, MalRAG first performs feature normalization consistent with the process described in Section~\ref{sec:database_construction}. Each available traffic view is converted into its normalized vector form before retrieval. Based on these representations, the adaptive retrieval module independently searches across the corresponding feature spaces to identify similar samples and their associated labels.
\par To improve retrieval precision, MalRAG first restricts the search space according to the protocol information associated with each feature view $m$. Specifically, before calculating similarities, the system filters the traffic database to include only flows that share the same protocol as the query flow. If the database contains flows with an identical \textit{fine-grained protocol label} (e.g., \texttt{TCP|TLS1.2}), the retrieval is performed within this subset; otherwise, it falls back to a \textit{coarser protocol category} (e.g., \texttt{TCP}). This hierarchical protocol filtering ensures that retrieved samples come from comparable communication contexts, reducing cross-protocol noise and improving the relevance of retrieved evidence. Subsequent distance computation and top-$k$ retrieval for view $m$ are performed over $\mathcal{D}^{(m)}_{\text{proto}}(x_q)$. Based on the filtered subset, the distance between the query flow $x_q^{(m)}$ and each database entry $x_i^{(m)}$ is computed using a metric tailored to the characteristics of that feature view.

For normalized payload vector, MalRAG employs the Hamming distance to measure symbol-wise differences:
\begin{equation}
\small
d_H(x_q^{(\text{payload})}, x_i^{(\text{payload})}) 
= \frac{1}{L} \sum_{j=1}^{L} \mathbb{I}\big(x_{q,j}^{(\text{payload})} \neq x_{i,j}^{(\text{payload})}\big),
\end{equation}
where $L$ is the payload vector length and $\mathbb{I}(\cdot)$ is the indicator function. 

For normalized packet-length and inter-arrival-time sequences, the Euclidean distance is used to measure spectrum-level similarity:
\begin{equation}
\small
d_E(x_q^{(m)}, x_i^{(m)}) = \|x_q^{(m)} - x_i^{(m)}\|_2, 
\quad m \in \{\text{length}, \text{time}\}.
\end{equation}

The system then retrieves the top-$k$ nearest traffic samples (i.e., those with the smallest distances) to the query flow for each view from the protocol-filtered database $\mathcal{D}_{\text{proto}}$:
\begin{equation}
\small
\mathcal{R}^{(m)} = 
\text{Top-}k_{\text{min}}\big(d(x_q^{(m)}, x_i^{(m)})\big), 
\quad x_i^{(m)} \in \mathcal{D}_{\text{proto}}^{(m)}.
\end{equation}
\par These view-specific retrieval sets are then merged into a unified evidence pool that captures complementary aspects of traffic behavior. Through Coverage-Enhanced Retrieval, this process improves the contextual consistency of the evidence and provides a solid foundation for the subsequent Traffic-Aware Adaptive Pruning stage.

\subsubsection{Traffic-Aware Adaptive Pruning}
After Coverage-Enhanced Retrieval, the initial candidate set $\mathcal{R}^{(m)}$ may still contain irrelevant or misleading evidence. To improve reliability, MalRAG prunes candidates using traffic-aware, class-wise distance thresholds. Since traffic characteristics (e.g., packet-length or timing patterns) vary significantly across protocols, these thresholds are estimated independently within each protocol group.

To obtain these thresholds, we estimate an intra-class distance distribution for each class–protocol pair. Specifically, for each class $c \in \text{LABEL\_SET}$ under protocol $p$ and traffic feature view $m$, the intra-class mean distance and standard deviation are computed as:
\begin{equation}
\small
\begin{aligned}
\bar{d}_{c,p}^{(m)} &= \frac{1}{|S_{c,p}|^2} 
\sum_{x_i^{(m)}, x_j^{(m)} \in S_{c,p}} d(x_i^{(m)}, x_j^{(m)}), \\[4pt]
\sigma_{c,p}^{(m)} &= \sqrt{\frac{1}{|S_{c,p}|^2} 
\sum_{x_i^{(m)}, x_j^{(m)} \in S_{c,p}} 
\big(d(x_i^{(m)}, x_j^{(m)}) - \bar{d}_{c,p}^{(m)}\big)^2 }.
\end{aligned}
\end{equation}
The corresponding refinement threshold is then defined as
\begin{equation}
\small
\tau_{c,p}^{(m)} = \bar{d}_{c,p}^{(m)} + \alpha \, \sigma_{c,p}^{(m)},
\end{equation}
where $\alpha$ controls the tolerance to intra-class variation within each protocol group.

During refinement, each retrieved sample $x_i^{(m)} \in \mathcal{R}^{(m)}$ is evaluated using the threshold $\tau_{c_i,p_i}^{(m)}$ of its associated class–protocol pair. Samples whose distances to the query exceed this limit are filtered out:
\begin{equation}
\small
\mathcal{R}_{\text{refined}}^{(m)} = 
\{\,x_i^{(m)} \in \mathcal{R}^{(m)} \mid d(x_q^{(m)}, x_i^{(m)}) \le \tau_{c_i,p_i}^{(m)}\,\}.
\end{equation}
This refinement strategy ensures that the retrieved evidence remains consistent with the inherent characteristics of the target traffic, improving retrieval precision and stability.

\subsection{Answer Generation}
\begin{figure}
    \centering
    \includegraphics[width=0.48\textwidth]{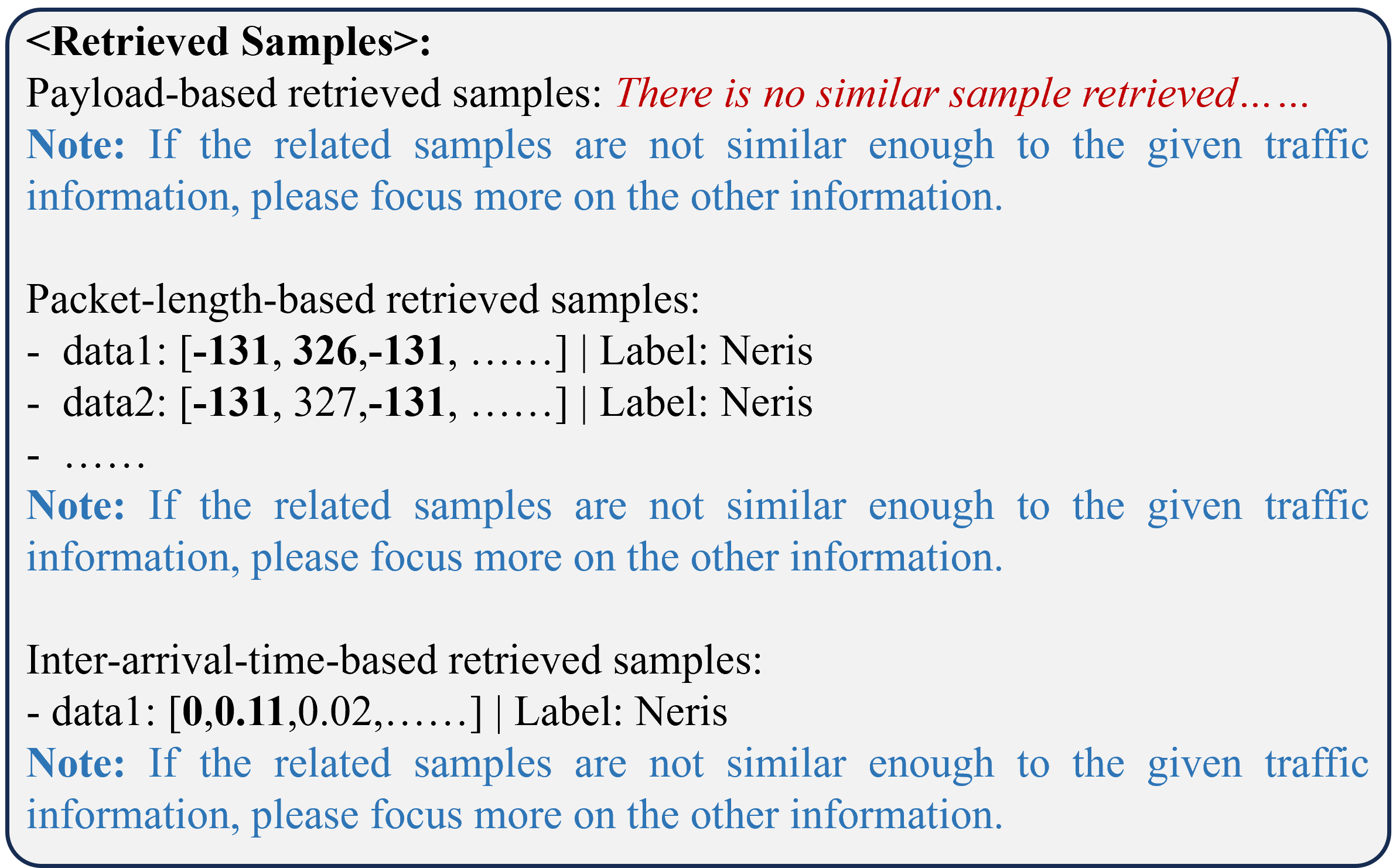}
    \caption{Example of the retrieved-evidence segment in the prompt.}
    \label{fig:full_prompt}
\end{figure}

After the prompt formulation and adaptive retrieval stages, MalRAG assembles the final query for the LLM by concatenating the guidance part with the retrieved-evidence segment. For each traffic feature view, this module first checks whether this view is available in the traffic database. If the view is available, the adaptive retrieval module selects a variable-size set of reliable evidence whose similarity passes a view-specific threshold, and inserts these samples together with their labels into the corresponding block. If the view exists in the database but no candidate passes the reliability test, a placeholder message is inserted such as \emph{``There are no similar samples retrieved for this view; please focus on other available information.''}. These conventions help the large language model adjust its reasoning when certain contextual evidence is missing or unreliable. Figure~\ref{fig:full_prompt} illustrates the retrieved-evidence segment under such mixed conditions: no payload-based samples are available, while the packet-length and inter-arrival-time views still provide usable retrieved evidence with different retained sample counts due to adaptive pruning. 

The completed prompt is then forwarded to the large language model, which produces the final output according to the user-specified \textit{reasoning option}. When reasoning mode is enabled, the model outputs both its analytical reasoning and the final decision, enabling human-in-the-loop inspection; otherwise, only the predicted label is returned for streamlined automated classification. 

\section{Performance Evaluation}
\subsection{Experimental Setup}
\subsubsection{Datasets Preprocessing}
\label{Datasets}
\par To avoid potential bias from dataset-specific identifiers, we adopt the randomization strategy described in~\cite{wickramasinghe2025sok}. Specifically, all datasets are preprocessed to randomize strong features such as IP addresses, port numbers, TCP sequence numbers, and TLS SNI fields, ensuring that model evaluation reflects genuine behavioral and structural learning rather than reliance on spurious artifacts.

\subsubsection{Implementation}
All experiments are conducted on an Ubuntu~22.04 server with four NVIDIA A800 GPUs (80~GB each), 1~TB RAM, and AMD EPYC~9654 CPUs. MalRAG is implemented in Python~3.10 with PyTorch~2.3.0, and the multi-view traffic database is managed in MongoDB for efficient storage and retrieval. The large language model backbone is \textbf{Qwen3-32B}, served via the vLLM inference framework in half-precision (FP16) mode for efficiency and stability. In Coverage-Enhanced Retrieval, the number of initially retrieved neighbors is fixed to $k=5$ per view. All modules are integrated into a unified pipeline to ensure consistent context handling and reproducibility.

\begin{table*}[htbp]
\caption{Comparison of known malicious traffic identification performance on different datasets \label{tab:known_results}}
\begin{center}
\setlength{\cmidrulekern}{0pt}
\renewcommand{\arraystretch}{1.6}
\resizebox{1\textwidth}{!}{
\begin{tabular}{
>{\arraybackslash}m{2.2cm}||
>{\centering\arraybackslash}m{1.4cm}>{\centering\arraybackslash}m{1.4cm}>{\centering\arraybackslash}m{1.6cm}|
>{\centering\arraybackslash}m{1.4cm}>{\centering\arraybackslash}m{1.4cm}>{\centering\arraybackslash}m{1.6cm}|
>{\centering\arraybackslash}m{1.4cm}>{\centering\arraybackslash}m{1.4cm}>{\centering\arraybackslash}m{1.6cm}}
\toprule
\multirow{2}{*}{\textbf{Method}} &
\multicolumn{3}{@{}c@{}|}{\textbf{CTU-13}} &
\multicolumn{3}{@{}c@{}|}{\textbf{USTC-2016}} &
\multicolumn{3}{@{}c@{}}{\textbf{DAPT-2020}} \\

\cmidrule(lr){2-4}\cmidrule(lr){5-7}\cmidrule(lr){8-10}
~ & \textbf{PRE} & \textbf{RCL} & \textbf{F1} & \textbf{PRE} & \textbf{RCL} & \textbf{F1} & \textbf{PRE} & \textbf{RCL} & \textbf{F1} \\
\midrule
APPScanner\cite{taylor2016appscanner}& 0.9079\textcolor{darkred}{$\blacktriangledown$4.3\%} & 0.6028\textcolor{darkred}{$\blacktriangledown$35.4\%} & 0.6645\textcolor{darkred}{$\blacktriangledown$28.7\%} &
0.7814\textcolor{darkred}{$\blacktriangledown$21.4\%} & 0.7575\textcolor{darkred}{$\blacktriangledown$23.6\%} & 0.7182\textcolor{darkred}{$\blacktriangledown$27.7\%} &
0.7590\textcolor{darkred}{$\blacktriangledown$20.0\%} & 0.7226\textcolor{darkred}{$\blacktriangledown$23.5\%} & 0.7408\textcolor{darkred}{$\blacktriangledown$21.6\%} \\

DF\cite{sirinam2018deep} & 0.6295\textcolor{darkred}{$\blacktriangledown$33.7\%} & 0.6449\textcolor{darkred}{$\blacktriangledown$30.9\%} & 0.6280\textcolor{darkred}{$\blacktriangledown$32.6\%} &
0.7600\textcolor{darkred}{$\blacktriangledown$23.6\%} & 0.6981\textcolor{darkred}{$\blacktriangledown$29.6\%} & 0.7109\textcolor{darkred}{$\blacktriangledown$28.4\%} &
0.7892\textcolor{darkred}{$\blacktriangledown$16.8\%} & 0.7759\textcolor{darkred}{$\blacktriangledown$17.8\%} & 0.7805\textcolor{darkred}{$\blacktriangledown$17.4\%} \\

FS-Net\cite{liu2019fs} & 0.7739\textcolor{darkred}{$\blacktriangledown$18.4\%} & 0.7187\textcolor{darkred}{$\blacktriangledown$23.0\%} & 0.7153\textcolor{darkred}{$\blacktriangledown$23.3\%} &
0.5964\textcolor{darkred}{$\blacktriangledown$40.0\%} & 0.7174\textcolor{darkred}{$\blacktriangledown$27.6\%} & 0.6371\textcolor{darkred}{$\blacktriangledown$35.8\%} &
0.8056\textcolor{darkred}{$\blacktriangledown$15.1\%} & 0.7783\textcolor{darkred}{$\blacktriangledown$17.6\%} & 0.7946\textcolor{darkred}{$\blacktriangledown$15.9\%} \\

AN-Net\cite{deng2024net} & 0.8758\textcolor{darkred}{$\blacktriangledown$7.7\%} & 0.8904\textcolor{darkred}{$\blacktriangledown$4.6\%} & 0.8783\textcolor{darkred}{$\blacktriangledown$5.8\%} &
0.9244\textcolor{darkred}{$\blacktriangledown$7.0\%} & 0.9195\textcolor{darkred}{$\blacktriangledown$7.3\%} & 0.9202\textcolor{darkred}{$\blacktriangledown$7.3\%} &
0.8967\textcolor{darkred}{$\blacktriangledown$5.5\%} & 0.9229\textcolor{darkred}{$\blacktriangledown$2.3\%} & 0.9076\textcolor{darkred}{$\blacktriangledown$3.9\%} \\

ET-BERT\cite{lin2022bert} & 0.6094\textcolor{darkred}{$\blacktriangledown$35.8\%} & 0.6408\textcolor{darkred}{$\blacktriangledown$31.4\%} & 0.6040\textcolor{darkred}{$\blacktriangledown$35.2\%} &
0.9298\textcolor{darkred}{$\blacktriangledown$6.5\%} & 0.9369\textcolor{darkred}{$\blacktriangledown$5.5\%} & 0.9299\textcolor{darkred}{$\blacktriangledown$6.3\%} &
0.9312\textcolor{darkred}{$\blacktriangledown$1.9\%} & 0.8556\textcolor{darkred}{$\blacktriangledown$9.4\%} & 0.8918\textcolor{darkred}{$\blacktriangledown$5.6\%} \\

YaTC\cite{zhao2023yet} &0.7615\textcolor{darkred}{$\blacktriangledown$19.7\%} & 0.7519\textcolor{darkred}{$\blacktriangledown$19.5\%} & 0.7385\textcolor{darkred}{$\blacktriangledown$20.8\%} &
0.9577\textcolor{darkred}{$\blacktriangledown$3.7\%} & 0.9677\textcolor{darkred}{$\blacktriangledown$2.4\%} & 0.9627\textcolor{darkred}{$\blacktriangledown$3.0\%} &
0.9310\textcolor{darkred}{$\blacktriangledown$1.9\%} & 0.9432\textcolor{darkred}{$\blacktriangledown$0.1\%} & 0.9371\textcolor{darkred}{$\blacktriangledown$0.8\%} \\

TFE-GNN\cite{zhang2023tfe} & 0.8156\textcolor{darkred}{$\blacktriangledown$14.0\%} & 0.8234\textcolor{darkred}{$\blacktriangledown$11.8\%} & 0.8135\textcolor{darkred}{$\blacktriangledown$12.7\%} &
0.9633\textcolor{darkred}{$\blacktriangledown$3.1\%} & 0.9656\textcolor{darkred}{$\blacktriangledown$2.6\%} & 0.9640\textcolor{darkred}{$\blacktriangledown$2.9\%} &
0.8350\textcolor{darkred}{$\blacktriangledown$12.0\%} & 0.8342\textcolor{darkred}{$\blacktriangledown$11.7\%} & 0.8345\textcolor{darkred}{$\blacktriangledown$11.7\%} \\

TrafficFormer\cite{zhou2025trafficformer} & 0.8364\textcolor{darkred}{$\blacktriangledown$11.8\%} & 0.8135\textcolor{darkred}{$\blacktriangledown$12.9\%} & 0.8238\textcolor{darkred}{$\blacktriangledown$11.6\%} &
0.9819\textcolor{darkred}{$\blacktriangledown$1.3\%} & 0.9795\textcolor{darkred}{$\blacktriangledown$1.2\%} & 0.9795\textcolor{darkred}{$\blacktriangledown$1.3\%} &
0.9365\textcolor{darkred}{$\blacktriangledown$1.30\%} & 0.9281\textcolor{darkred}{$\blacktriangledown$1.70\%} & 0.9301\textcolor{darkred}{$\blacktriangledown$1.60\%} \\

\textbf{MalRAG} &\textbf{0.9488}\textcolor{darkgreen}{(base)} & \textbf{0.9336}\textcolor{darkgreen}{(base)} & \textbf{0.9320}\textcolor{darkgreen}{(base)} &
\textbf{0.9944}\textcolor{darkgreen}{(base)} & \textbf{0.9915}\textcolor{darkgreen}{(base)} & \textbf{0.9928}\textcolor{darkgreen}{(base)} &
\textbf{0.9491}\textcolor{darkgreen}{(base)} & \textbf{0.9443}\textcolor{darkgreen}{(base)} & \textbf{0.9449}\textcolor{darkgreen}{(base)} \\
\bottomrule
\end{tabular}
}
\end{center}
\begin{flushleft}
\footnotesize \textcolor{darkred}{$\blacktriangledown$} denotes the percentage of decrease and improvement in the corresponding metric compared to the \textcolor{darkgreen}{base} (Ours).
\end{flushleft}
\end{table*}

\subsection{Known Malicious Traffic Identification}
\subsubsection{Experimental Settings}
We evaluate MalRAG on CTU-13, DAPT-2020, and USTC-2016 for the known malicious traffic identification task. For each dataset, we partition flows at the session level to avoid packet-level leakage: 80\% of labeled flows are used to populate the MalRAG traffic database, and the remaining 20\% are held out as the test set. Splits are performed in a class- and protocol-stratified manner to preserve class balance and protocol diversity in both database and test subsets. All public datasets undergo the strong-feature randomization described in Section~\ref{Datasets} to remove spurious identifiers prior to database construction. To reduce variance due to a particular split, we repeat the stratified partitioning with five different random seeds and report the average results across these five runs.

When building the database, we index each stored flow by its flow ID, fine-grained protocol tag (e.g., \texttt{TCP|TLS1.2} when available), and available feature views. During testing, each query flow is processed as described in Section~4: the user-provided views (payload/length/time as available) are normalized and used to construct the prompt placeholders, Coverage-Enhanced Retrieval and Traffic-Aware Adaptive Pruning are executed, and the resulting prompt plus retrieved evidence is sent to the LLM for identification.

\subsubsection{Comparison Methods}
We compare MalRAG with a series of representative methods covering machine learning based, deep learning based, and pretrained based methods to comprehensively evaluate its capability in identifying known malicious traffic. The comparison methods include (i) machine learning based methods: AppScanner\cite{taylor2016appscanner}; (ii) deep learning based methods: DF\cite{sirinam2018deep}, FS-Net\cite{liu2019fs}, AN-Net\cite{deng2024net}; (iii) pretrained based methods: ET-BERT\cite{lin2022bert}, YaTC\cite{zhao2023yet}, TFE-GNN\cite{zhang2023tfe}, TrafficFormer\cite{zhou2025trafficformer}.

\subsubsection{Evaluation Metrics}
For the known malicious traffic identification task, we evaluate performance using standard classification metrics, including precision (PRE), recall (RCL), and F1-score (F1). Each malicious class is treated as the positive class while all others are considered negative when computing these metrics. To account for class imbalance across different malicious classes, we report the macro-averaged precision, recall, and F1-score over all known malicious categories, providing a balanced and comprehensive assessment of identification accuracy.

\subsubsection{Evaluation Results}
\begin{figure}[htbp]
\centering
  \subfloat[TrafficFormer]{\includegraphics[width=0.45\linewidth]{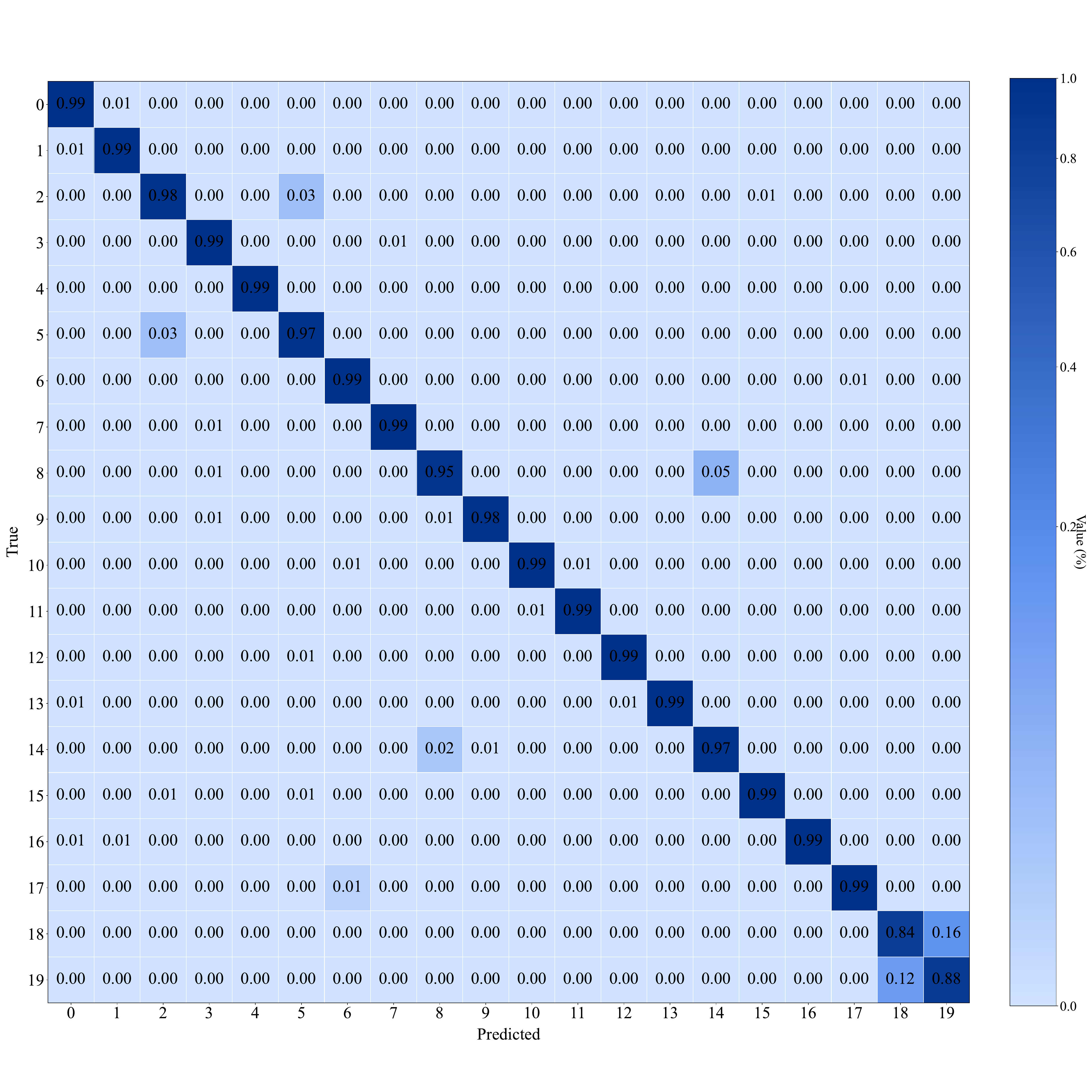}}
  \subfloat[MalRAG]{\includegraphics[width=0.45\linewidth]{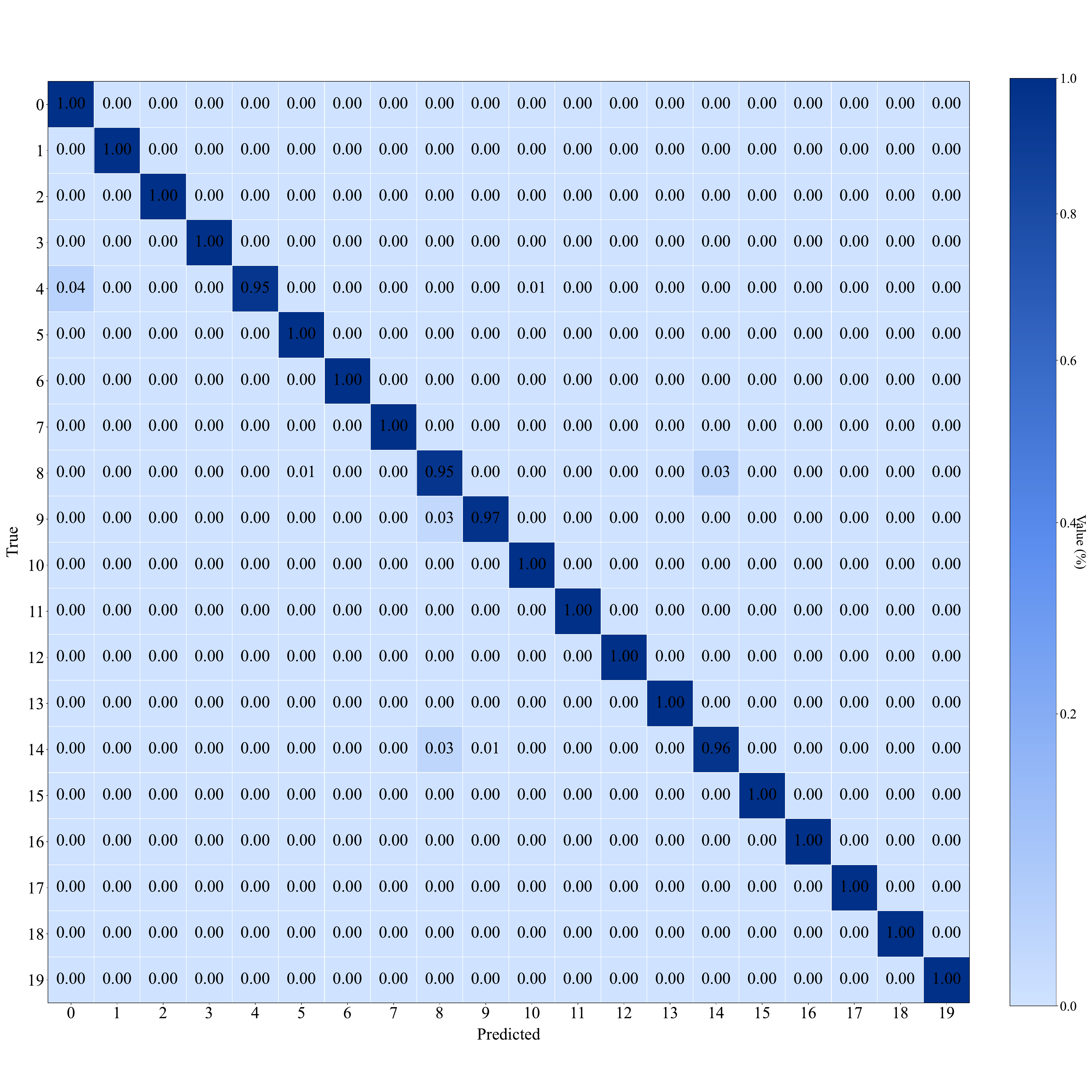}}\hfill\\
  \caption{Confusion matrix of each method on USTC-2016 dataset. Class indices 0–19 correspond to BitTorrent, Cridex, Facetime, FTP, Geodo, Gmail, Htbot, Miuref, MySQL, Neris, Nsis-ay, Outlook, Shifu, Skype, SMB, Tinba, Virut, Weibo, WorldOfWarcraft, and Zeus, respectively.}
  \label{fig:misclassification_ustc-2016}
\end{figure}

\begin{figure}[htbp]
\centering
  \subfloat[YaTC]{\includegraphics[width=0.3\linewidth]{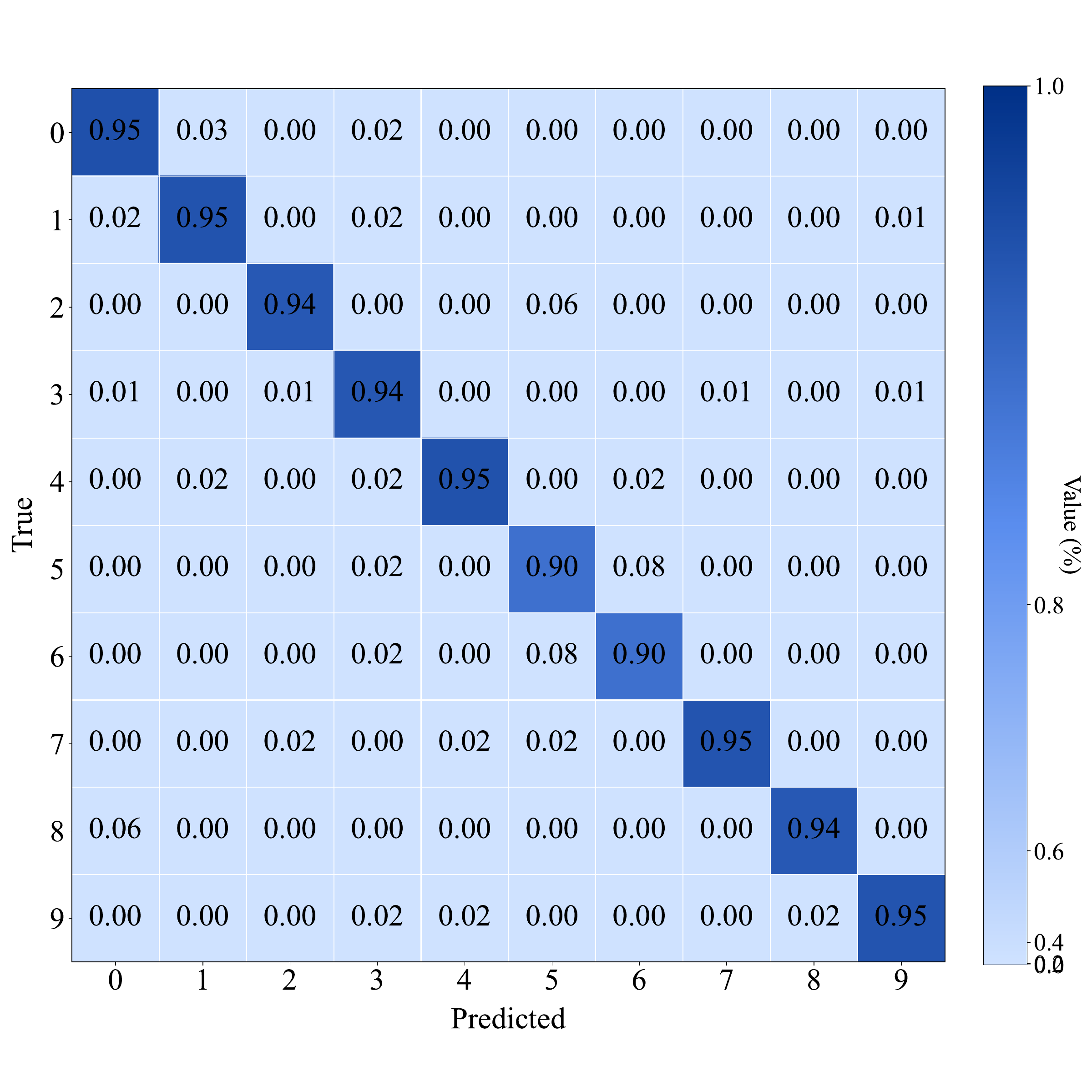}}
  \subfloat[TrafficFormer]{\includegraphics[width=0.3\linewidth]{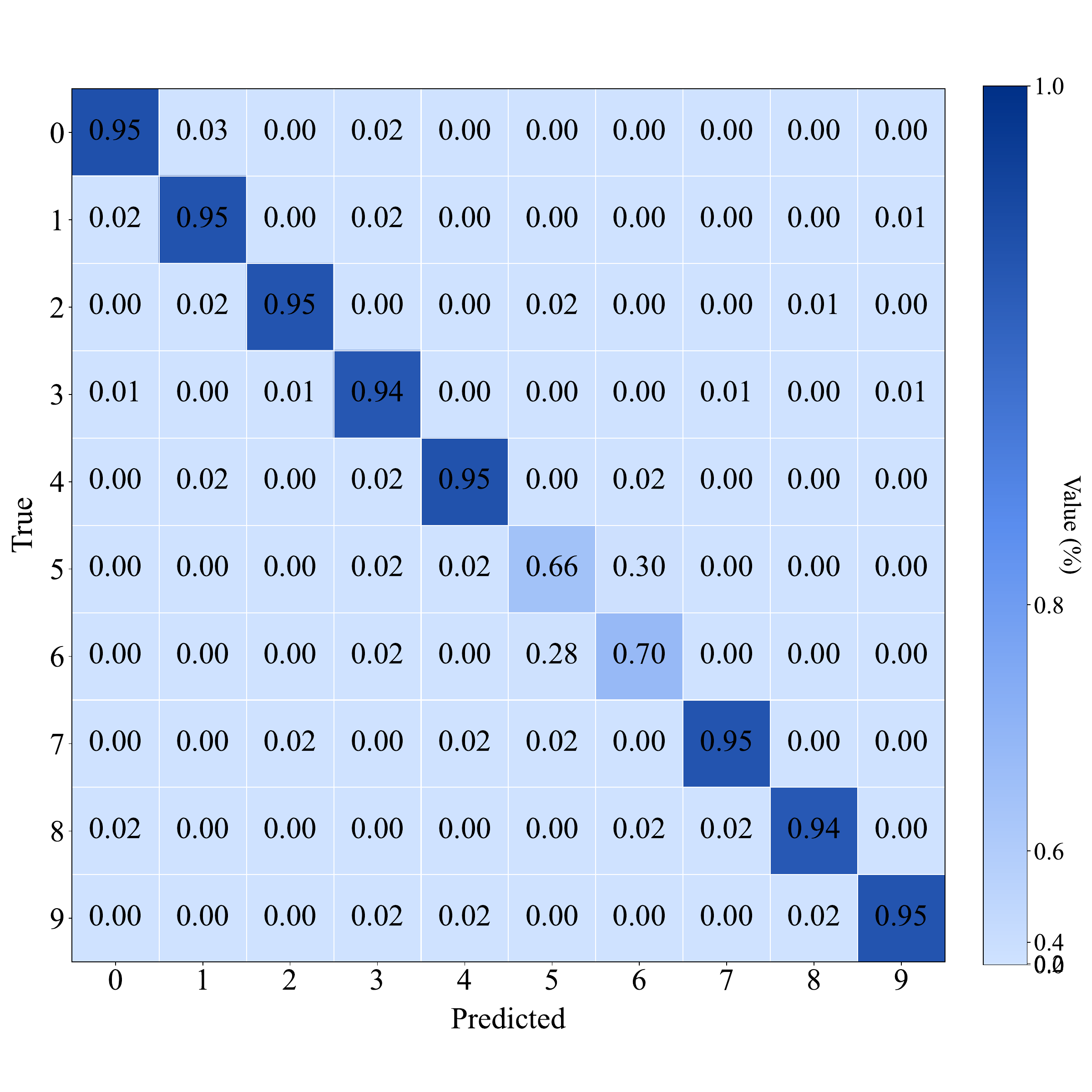}}
  \subfloat[MalRAG]{\includegraphics[width=0.3\linewidth]{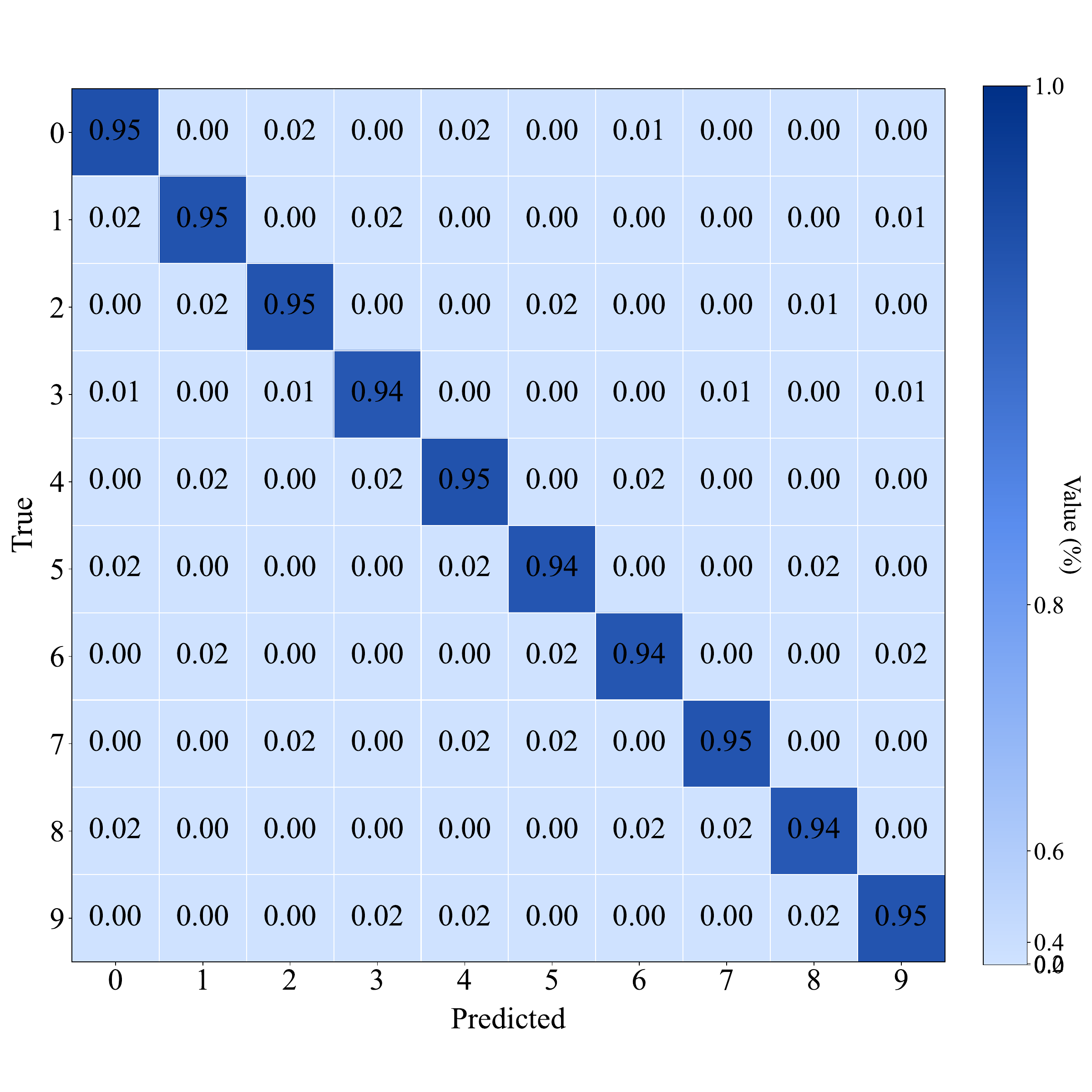}}\hfill\\
  \caption{Confusion matrix of each method on DAPT-2020 dataset. Class indices 0–9 correspond to Account bruteforce, Backdoor, Command Injection, CSRF, DoS, Malware Download, Network Scan, Privilege escalation, SQL injection, and Web Vulnerability Scan, respectively.}
  \label{fig:misclassification_dapt-2020}
\end{figure}

Table~\ref{tab:known_results} presents the quantitative results for known malicious traffic identification. Figure \ref{fig:misclassification_CTU-13},\ref{fig:misclassification_ustc-2016} and \ref{fig:misclassification_dapt-2020} show the confusion matrix of each methods. Based on the results, we summarize the following key observation:

\begin{figure}[htbp]
\centering
  \subfloat[YaTC]{\includegraphics[width=0.3\linewidth]{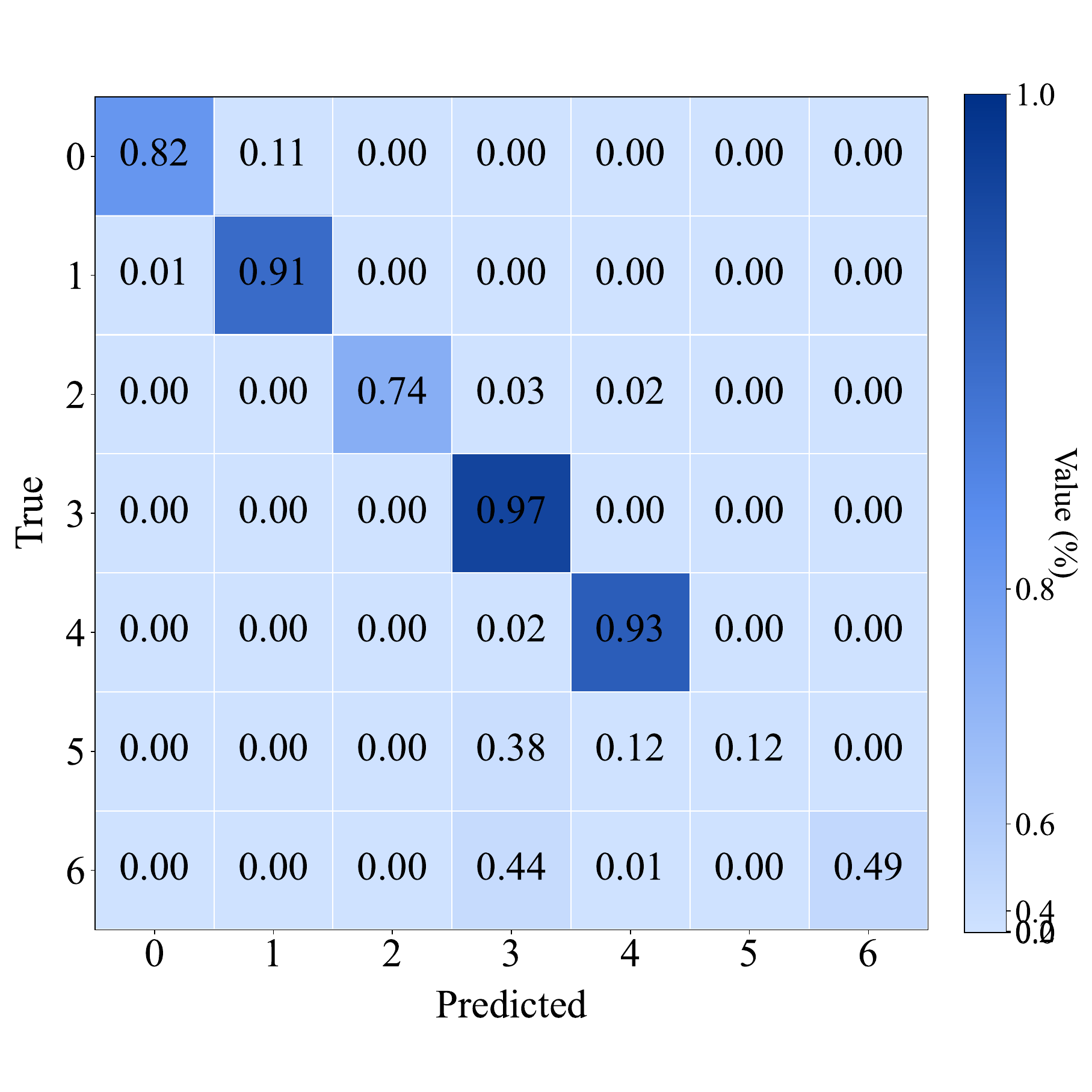}}
  \subfloat[TrafficFormer]{\includegraphics[width=0.3\linewidth]{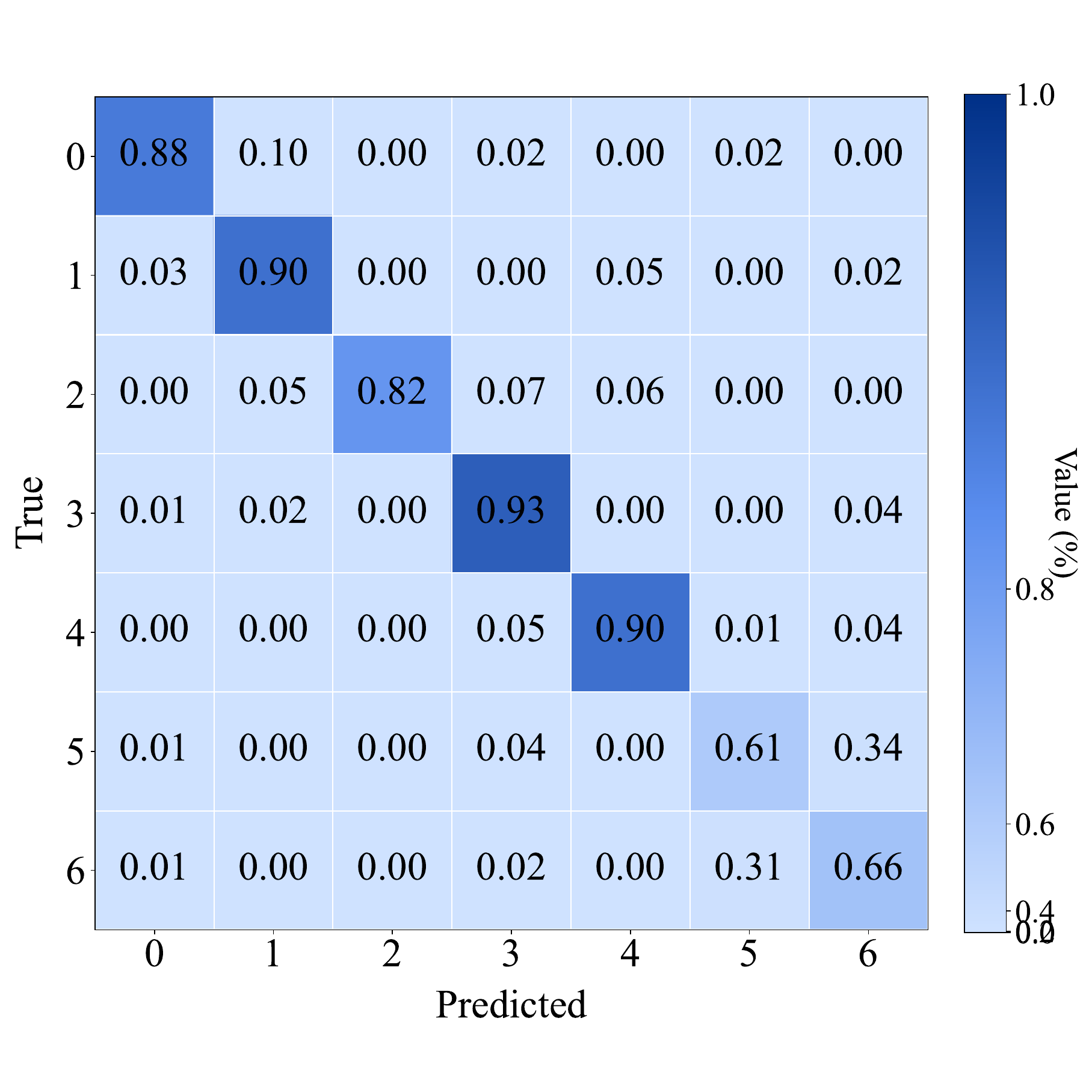}}
  \subfloat[MalRAG]{\includegraphics[width=0.3\linewidth]{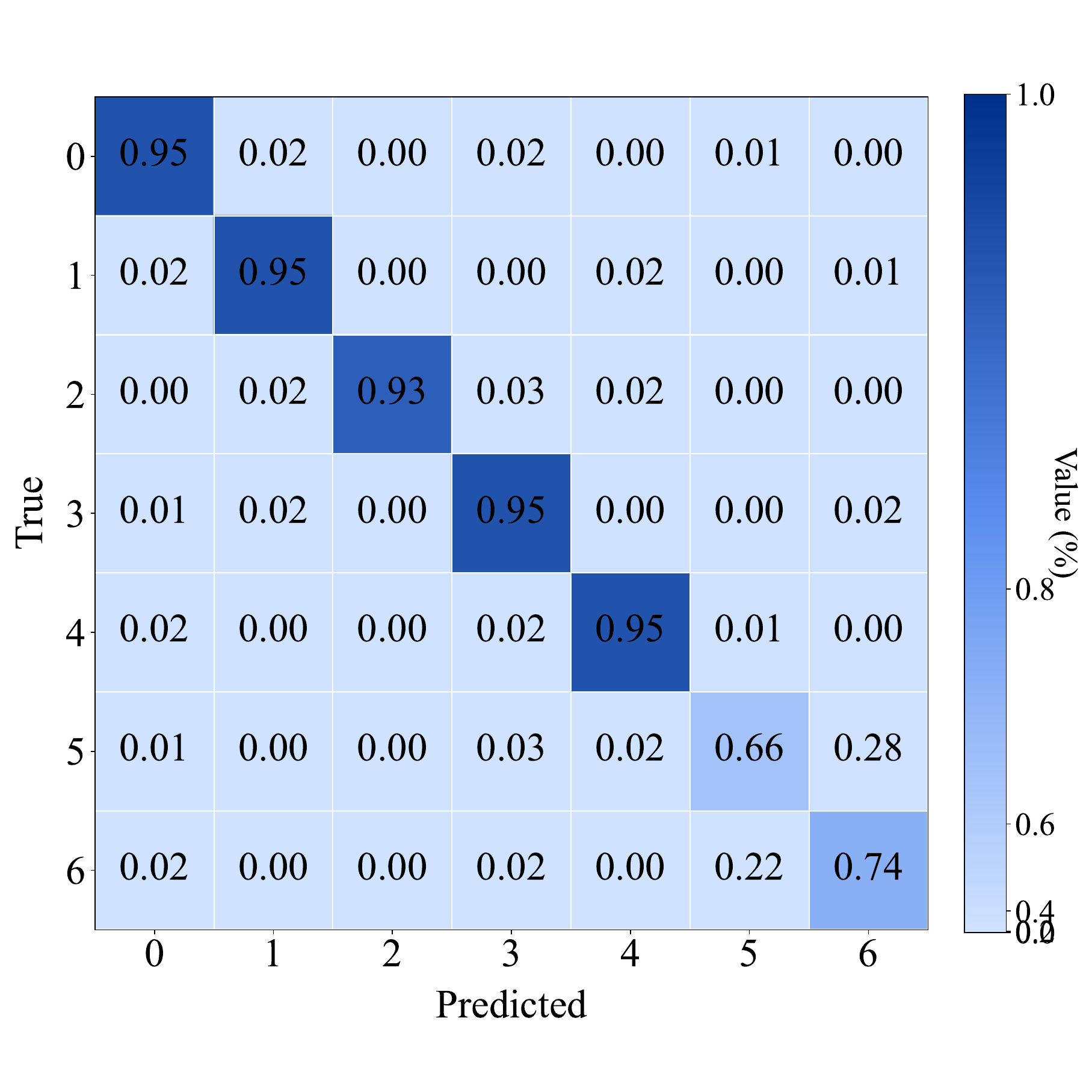}}\hfill\\
  \caption{Confusion matrix of each method on CTU-13 dataset. Class indices 0–6 correspond to Neris, Rbot, Virut, Menti, Sogou, Murlo, and NSIS.ay.}
  \label{fig:misclassification_CTU-13}
\end{figure}

\par\textbf{(1) MalRAG achieves state-of-the-art performance in malicious traffic identification across all evaluation datasets.} As shown in Table~\ref{tab:known_results}, MalRAG consistently outperforms comparison baselines in precision, recall, and F1-score on datasets like CTU-13, USTC-2016, and DAPT-2020. This advantage holds across different network environments and traffic types, demonstrating MalRAG's ability to accurately identify malicious behaviors in diverse scenarios. This superior performance can be attributed to the overall framework design, where the LLM is guided by structured prompts and grounded in retrieved traffic evidence. By combining linguistic reasoning with traffic-domain knowledge, MalRAG is able to make more nuanced, context-aware judgments than traditional methods.

\par\textbf{(2) MalRAG demonstrates superior cross-dataset generalization and task comprehension.} As shown in Table~\ref{tab:known_results}, MalRAG consistently outperforms other methods across different datasets. This superior performance is largely due to the framework’s ability to understand task requirements and context through the structured prompt design. By guiding the LLM with task instructions and relevant exemplar references, MalRAG achieves better generalization across diverse datasets compared to other methods, showing more robust performance without the need for fine-tuning.

\par\textbf{(3) MalRAG achieves high precision in malicious traffic identification due to its Coverage-Enhanced Retrieval mechanism.} By assessing suspicious flows from multiple complementary views, such as content, structural, and temporal characteristics, MalRAG retrieves highly relevant and reliable evidence. This multi-dimensional approach improves the precision by capturing diverse aspects of the traffic behavior, making the identification process more trustworthy. 

\par\textbf{(4) Retrieval-guided evidence reduces misclassification and sharpens class boundaries.}
As shown in Figure \ref{fig:misclassification_CTU-13}, \ref{fig:misclassification_ustc-2016}, and \ref{fig:misclassification_dapt-2020}, MalRAG achieves fewer misclassifications compared to the top-performing methods. This is due to its retrieval-guided evidence mechanism. By performing multi-dimensional retrieval and applying Traffic-Aware Adaptive Pruning, MalRAG ensures that only the most relevant and consistent samples are considered for classification. This refinement process filters out irrelevant evidence, allowing MalRAG to better differentiate between classes and significantly reduce false positives. The overall result is sharper class boundaries, leading to more accurate and reliable identification of malicious traffic.

\par\textbf{(5) MalRAG adapts to diverse tasks without fine-tuning and delivers stable performance.}
Across heterogeneous datasets and traffic types, MalRAG maintains consistent precision and recall while requiring no additional training or parameter updates. By combining structured prompting with retrieval of semantically relevant exemplars, the model tailors its reasoning to each task and data regime rather than relying on memorized parameters. This training-free adaptation reduces operational cost and mitigates overfitting, yielding reliable decisions under new attack patterns.

\subsection{Novel Malicious Traffic Discovery}
\subsubsection{Experimental Settings}  
To evaluate MalRAG’s ability to discover novel malicious traffic, we adopt an open-set setting where previously unseen attacks appear at test time. The traffic database is built exclusively from the earlier CTU-13 corpus. We then test three chronological scenarios: CTU-13 mixed with DAPT-2020, AndroidMischiefDataset, and our self-collected Malicious\_c2 dataset (Quakbot\_C2, Gozi\_C2, Tofsee\_C2, Trickbot\_C2). All novel flows originate from corpora collected after CTU-13, enforcing a time-ordered split that mirrors the emergence of new attacks. In each scenario, CTU-13 classes are treated as known and the additional traffic as novel. Preprocessing and database construction follow the same feature normalization as in Section~\ref{sec:database_construction}.
\par For comparison, all methods are trained exclusively on CTU-13 and evaluated under the chronological scenarios described above. We also run a separate experiment where their novelty detectors are removed and the architectures are assessed only on known malicious traffic identification on CTU-13. This setup reveals each architecture’s baseline known-class performance, and we report results both with and without the novelty detector.

\subsubsection{Comparison Methods}  
We compare MalRAG with the following five representative methods that can simultaneously identify known malicious traffic and perform novel malicious traffic discovery, including OpenMax~\cite{OpenMax}, ZTI~\cite{jin2020zero}, GMAF~\cite{xia2021gmaf}, CADE~\cite{yang2021cade}, and ICE-CP~\cite{luo2024identifying}.

\subsubsection{Evaluation Metrics}  
For this task, we employ five metrics that jointly assess known malicious traffic identification and novel malicious traffic discovery performance:  
(1) average precision of known classes ($\text{PRE-K}$),  
(2) average recall of known classes ($\text{RCL-K}$),  
(3) average precision of novel attacks ($\text{PRE-N}$),  
(4) average recall of novel attacks ($\text{RCL-N}$), and  
(5) Normalized Accuracy (NA) \cite{mendes2017nearest}, which captures the trade-off between known malicious traffic identification and novel malicious traffic discovery.  
In addition, we report the macro F1-score over known classes to facilitate comparison of known malicious traffic identification across different architectures.

\subsubsection{Evaluation Results}
\begin{figure}
    \centering
    \subfloat[]{\includegraphics[width=0.3\linewidth]{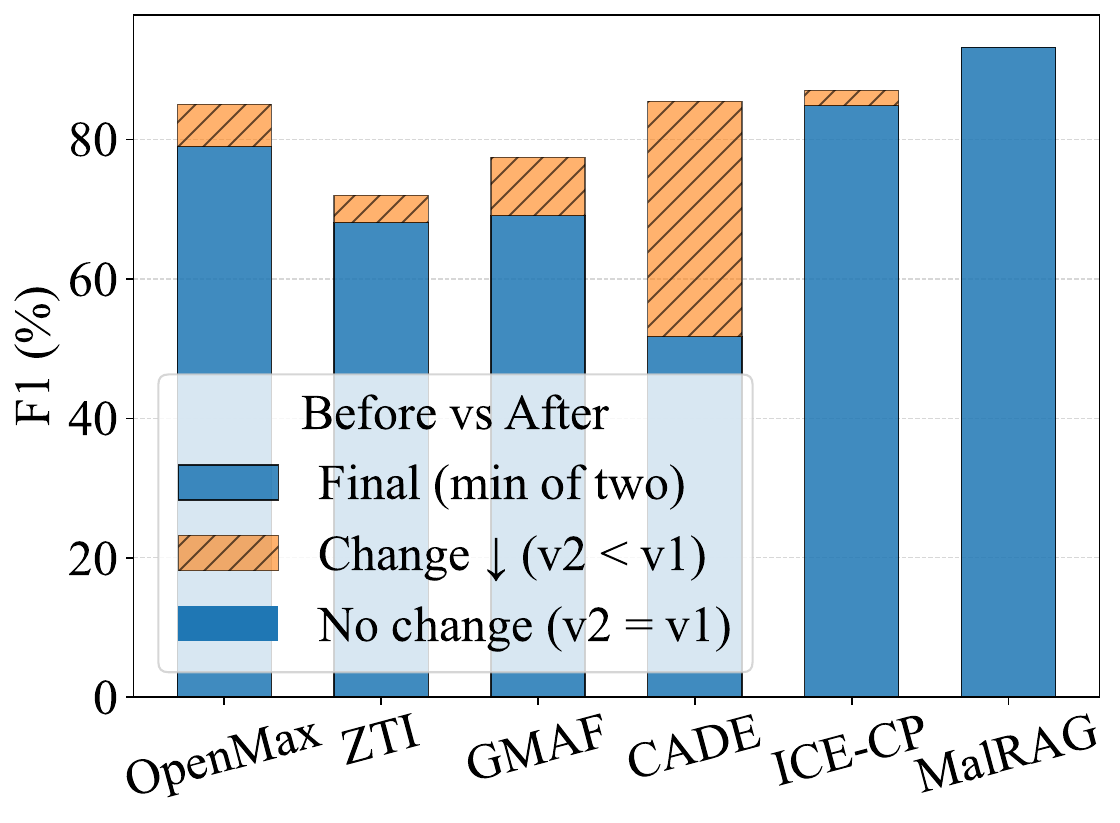}}
    \subfloat[]{\includegraphics[width=0.3\linewidth]{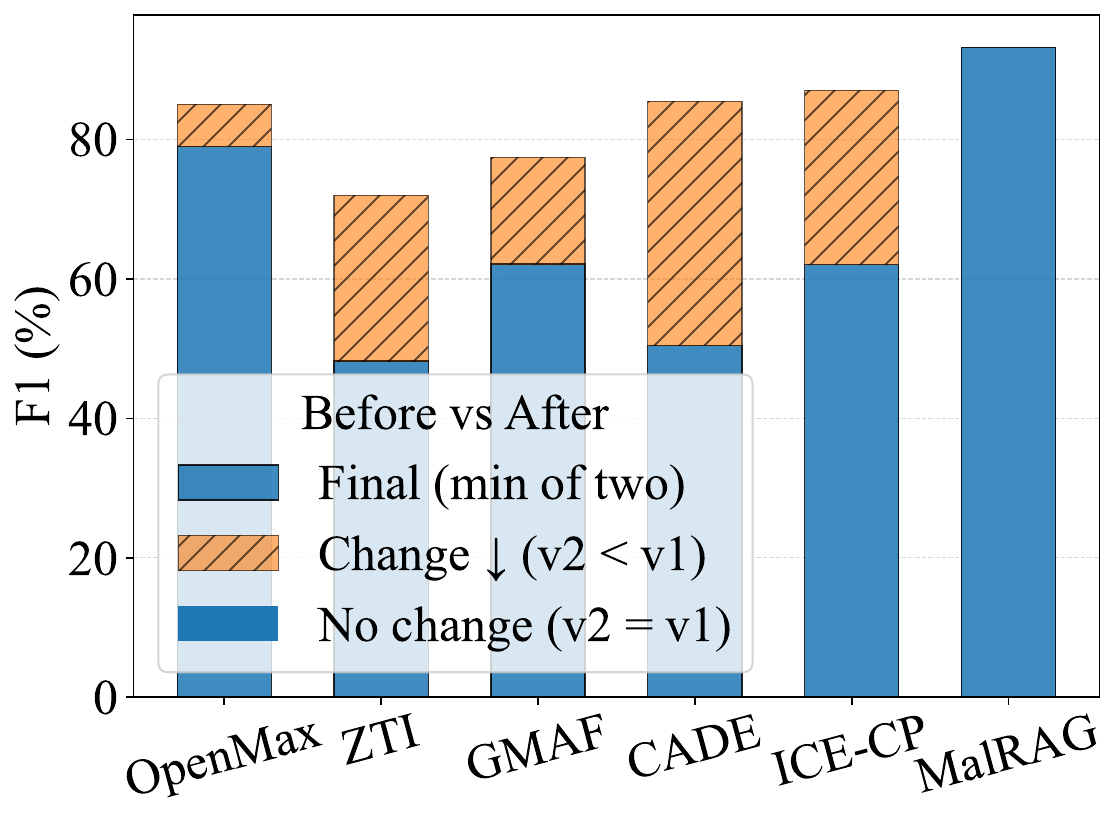}}
    \subfloat[]{\includegraphics[width=0.3\linewidth]{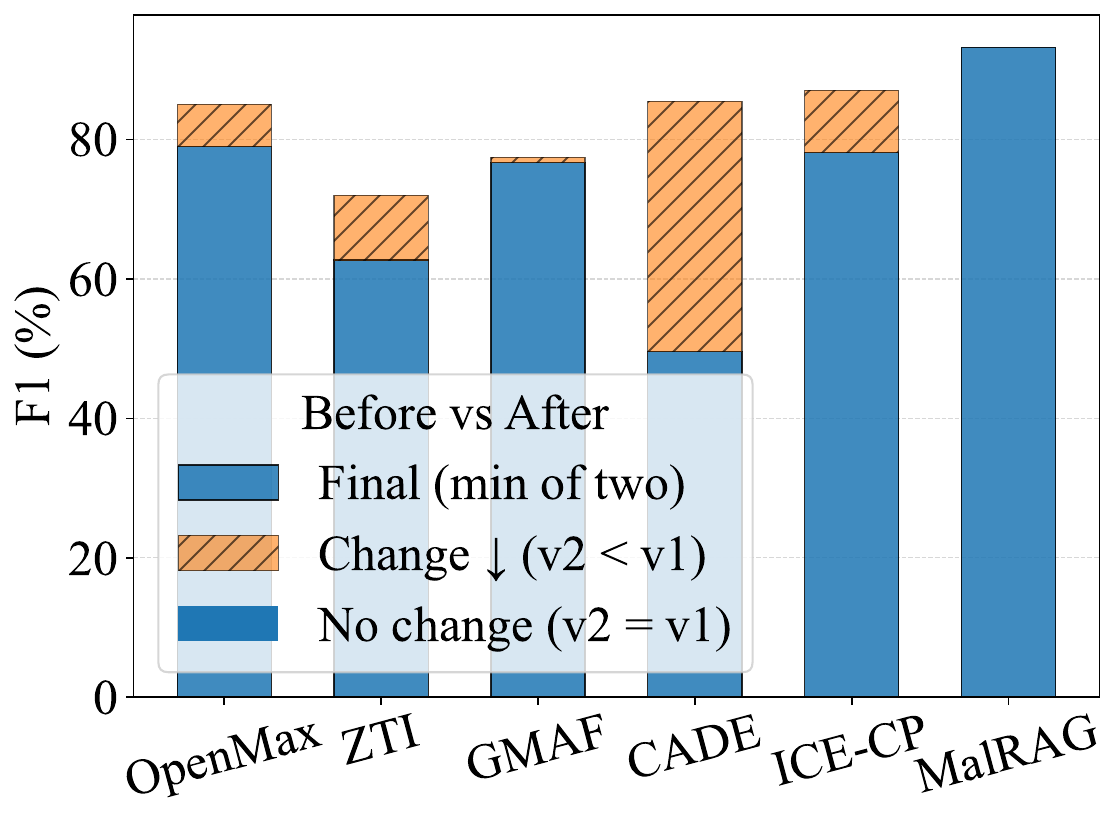}}
    \label{fig:delta}
    \caption{Impact of adding a novelty detector on known-class identification of each compared method. (a) Android Mischief dataset as novel malicious traffic, (b) DAPT-2020 dataset as novel malicious traffic, (c) Malicious\_c2 dataset as novel malicious traffic.}
\end{figure}

Figure~\ref{fig:novel_radar} presents the performance comparison of different methods across 3 evaluation scenarios for novel malicious traffic discovery. Figure \ref{fig:delta} shows how adding a novelty detector affects each compared method’s performance on known-class identification by contrasting results with and without the detector. Based on these results, we summarize the following key observations:
\begin{figure}
  \centering
  \subfloat[]{\includegraphics[width=0.3\linewidth]{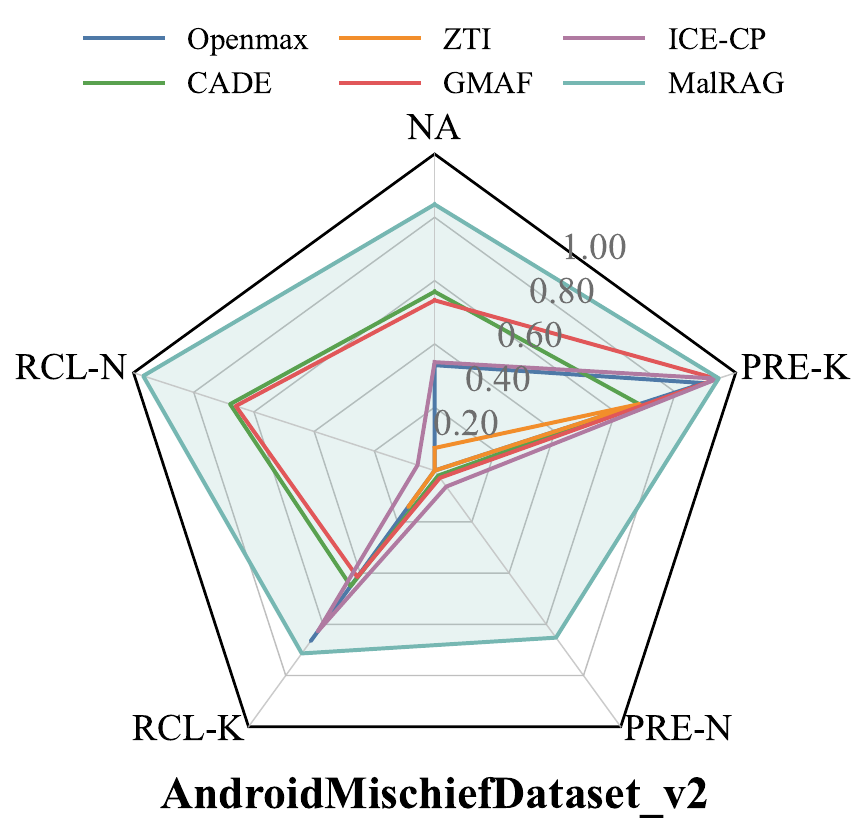}}
  \subfloat[]{\includegraphics[width=0.3\linewidth]{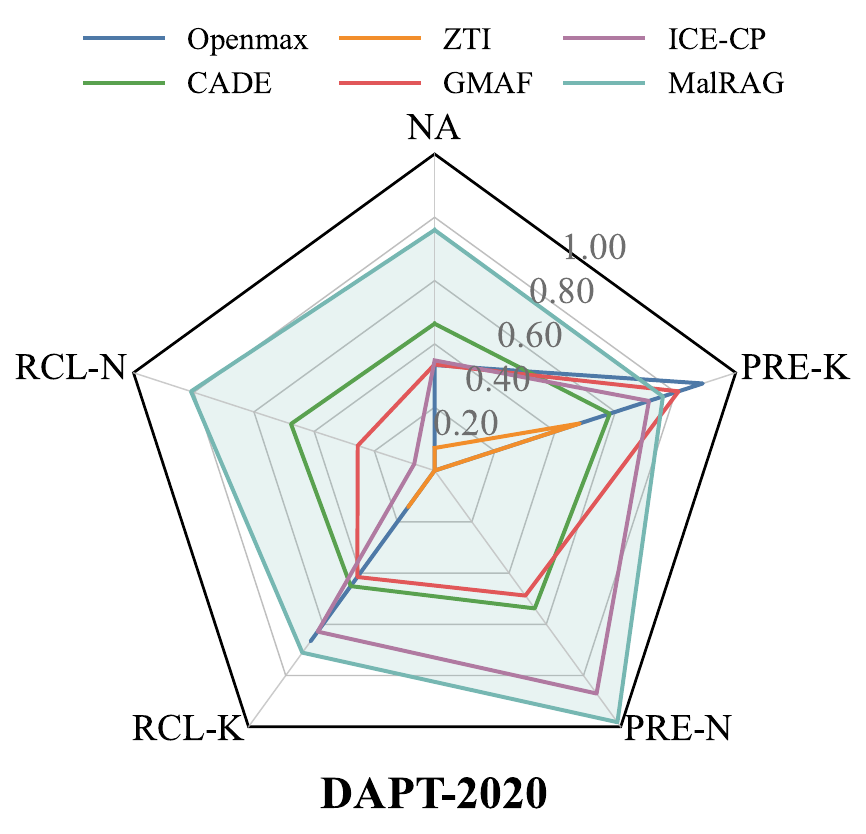}}
  \subfloat[]{\includegraphics[width=0.3\linewidth]{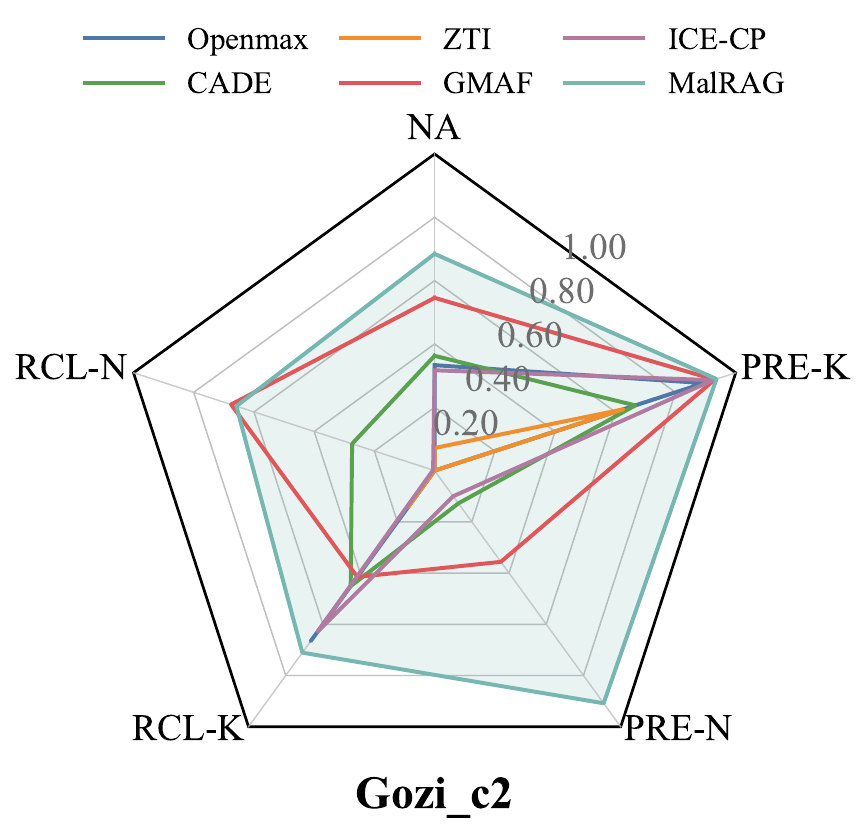}}\hfill
  \caption{Comparison of novel malicious traffic discovery performance on different datasets. (a) Android Mischief dataset as novel malicious traffic, (b) DAPT-2020 dataset as novel malicious traffic, (c) Malicious\_c2 dataset as novel malicious traffic.}\label{fig:novel_radar}
\end{figure}

\par\textbf{(1) MalRAG consistently achieves the highest performance across all evaluation scenarios.}
As illustrated in Figure~\ref{fig:novel_radar}, MalRAG outperforms all baseline methods in both known malicious traffic identification and novel malicious traffic discovery. It achieves the best precision and recall for both known and novel classes, as well as the highest NA across datasets, demonstrating its ability to maintain balanced detection under varying traffic compositions.

\par\textbf{(2) The retrieval-augmented design enables accurate discovery of novel malicious traffic without compromising known malicious traffic precision.}
While traditional approaches such as CADE and ICE-CP face a trade-off between novel detection sensitivity and known-class stability, MalRAG benefits from Coverage-Enhanced Retrieval that provides semantically relevant evidence.
This allows the LLM to distinguish genuine novel behaviors from normal variations, improving recall for novel malicious traffic while sustaining high precision for known classes.

\par\textbf{(3) MalRAG shows remarkable adaptability across heterogeneous domains and malicious traffic classes.}
Whether evaluated on Android application traffic, newly collected C2 families, or different public datasets, MalRAG maintains stable normalized accuracy between 0.72 and 0.85.
This robustness demonstrates its capability to generalize beyond specific traffic protocols and data collection environments, confirming the effectiveness of its protocol-aware retrieval and prompt-based reasoning mechanism.

\par\textbf{(4) MalRAG achieves high performance without additional model training or fine-tuning in novel malicious traffic discovery.}
Unlike learning-based open-world models that require retraining when new traffic appears, MalRAG performs dynamic reasoning through prompt construction and retrieval, making it immediately deployable in evolving network environments.
This training-free property highlights its practical advantage in post-detection analysis, where novel malicious traffic continuously emerges.

\subsection{Ablation Study}
To comprehensively assess the contribution of each core module in MalRAG, we conduct ablation experiments across three tasks: known malicious traffic identification and novel malicious traffic discovery.  
All experiments are performed under identical environments, using the same LLM backbone, traffic database, and prompt construction described in previous sections.  
The goal is to isolate and analyze the effect of each major component, including the \textbf{Coverage-Enhanced Retrieval (CER)}, \textbf{Traffic-Aware Adaptive Pruning (TAP)}, and \textbf{guidance prompt (GP)}.  
Four model variants are compared to quantify individual and combined effects:
\begin{itemize}
    \item \textbf{w/o CER:} The LLM directly analyzes the raw traffic input without retrieval, to quantify the benefit of incorporating comprehensive evidence.
    \item \textbf{w/o TAP:} CER is retained, but all top-$k$ retrieved samples are included in the prompt without pruning, to assess the importance of filtering low-quality evidence.
    \item \textbf{w/o GP:} Only the task instruction is kept while other guidance is removed, evaluating how the guidance prompt affects evidence reference and decision quality.
    \item \textbf{Full MalRAG:} The complete framework integrating CER, TAP, and GP.
\end{itemize}
This setup enables quantifying the contribution of contextual grounding, evidence filtering, and prompt-guided reasoning under consistent evaluation settings.

\begin{table}[]
\centering
\caption{Ablation results for known malicious traffic identification measured by F1.}
\begin{tabular}{lccc}
\toprule
\textbf{Method} & \textbf{DAPT-2020} & \textbf{CTU-13} & \textbf{USTC-2016} \\
\midrule
\textbf{Full (Ours)} & \textbf{0.9615} & \textbf{0.8941} & \textbf{0.9928} \\
w/o GP               & 0.9242 & 0.8607 & 0.9678 \\
w/o CER             & 0.3021 & 0.5017 & 0.0248 \\
w/o TARP               & 0.9443 & 0.8801 & 0.9838 \\
\bottomrule
\end{tabular}
\label{tab:ablation_known}
\end{table}

\subsubsection{Ablation Results for Known Malicious Traffic Identification}
\begin{table*}[t]
\centering
\footnotesize
\caption{Performance comparison of MalRAG with different backbone LLMs on known malicious traffic identification across three datasets.}
\label{tab:deepdive_backbone}
\begin{tabular}{c lccccccccc}
\toprule
\multirow{2}{*}{\textbf{No}} & \multirow{2}{*}{\textbf{Backbone LLM}} & 
\multicolumn{3}{c}{\textbf{USTC-2016}} & 
\multicolumn{3}{c}{\textbf{CTU-13}} & 
\multicolumn{3}{c}{\textbf{DAPT-2020}} \\
\cmidrule(lr){3-5} \cmidrule(lr){6-8} \cmidrule(lr){9-11}
 & & \textbf{PRE} & \textbf{RCL} & \textbf{F1} & 
   \textbf{PRE} & \textbf{RCL} & \textbf{F1} & 
   \textbf{PRE} & \textbf{RCL} & \textbf{F1} \\
\midrule
1 & \textbf{Qwen3-32B (ours)} & \textbf{0.9944} & \textbf{0.9915} & \textbf{0.9928} & \textbf{0.9488} & \textbf{0.9336} & \textbf{0.9320} & \textbf{0.9491} & \textbf{0.9443} & \textbf{0.9449} \\
2 & Qwen3-14B & 0.9927 & 0.9256 & 0.9530 & 0.9198 & 0.8861 & 0.8955 & 0.9344 & 0.9176 & 0.9234 \\
3 & Qwen3-8B & 0.9889 & 0.9326 & 0.9560 & 0.8738 & 0.8378 & 0.8525 & 0.8485 & 0.7270 & 0.7228 \\
4 & Qwen3-30B-A3B-Instruct-FP8 & 0.9782 & 0.9770 & 0.9770 & 0.8998 & 0.9046 & 0.8997 & 0.9003 & 0.8755 & 0.8735 \\
5 & Meta-Llama-3-8B & 0.9884 & 0.6499 & 0.7670 & 0.9044 & 0.5496 & 0.6817 & 0.9261 & 0.3794 & 0.5112 \\
6 & Mistral-7B-Instruct-v0.3 & 0.9765 & 0.7319 & 0.8231 & 0.9159 & 0.7451 & 0.8144 & 0.9373 & 0.7833 & 0.8509 \\
7 & ChatGLM4-9B & 0.9825 & 0.9367 & 0.9543 & 0.9094 & 0.8297 & 0.8114 & 0.7694 & 0.5676 & 0.4723 \\
8 & Yi-1.5-34B-Chat & 0.9817 & 0.8125 & 0.8844 & 0.8952 & 0.8641 & 0.8768 & 0.9365 & 0.9058 & 0.9192 \\
\bottomrule
\end{tabular}
\end{table*}

We first evaluate the contribution of each module to known malicious traffic identification, using the same database construction and data splits as in Section~\ref{Datasets}. Each ablation variant is independently assessed, and the results are summarized in Table~\ref{tab:ablation_known}. When retrieval is removed, the LLM is forced to analyze traffic in isolation without contextual grounding, which performs poorly given the complexity and similarity of encrypted traffic and highlights the need for a well-structured traffic database to supply representative evidence. In addition, both Traffic-Aware Adaptive Pruning and the guidance prompt prove important: the former filters irrelevant or misleading samples to improve reliability, while the latter provides analytical cues that lead to more stable and domain-aligned decisions.

\subsubsection{Ablation Results for Novel Malicious Traffic Discovery}
\begin{figure}[]
  \centering
  \subfloat[]{\includegraphics[width=0.3\linewidth]{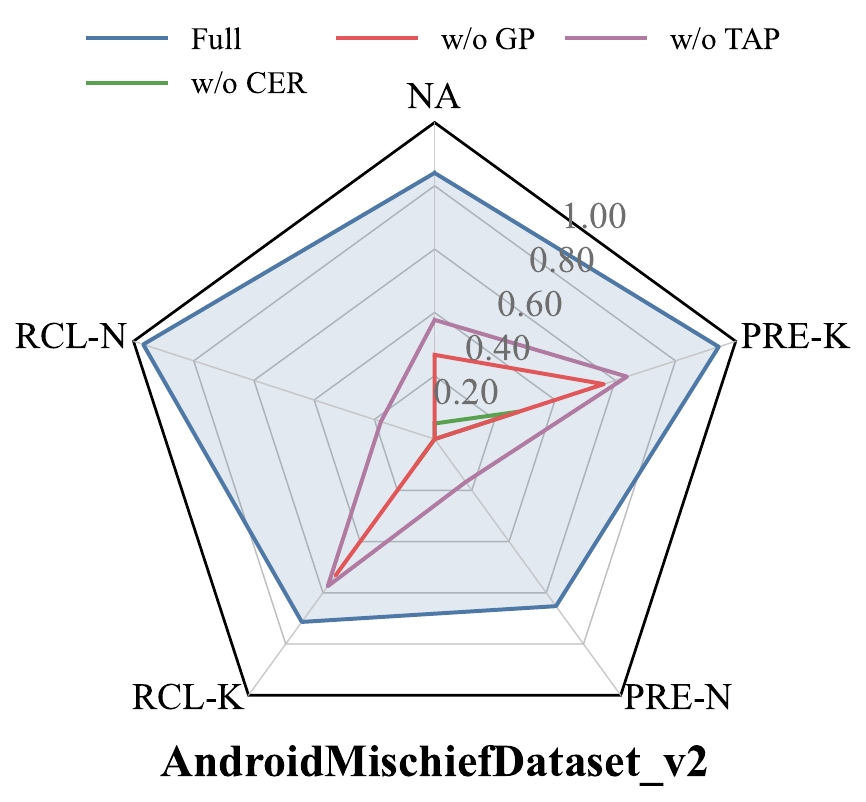}}
  \subfloat[]{\includegraphics[width=0.3\linewidth]{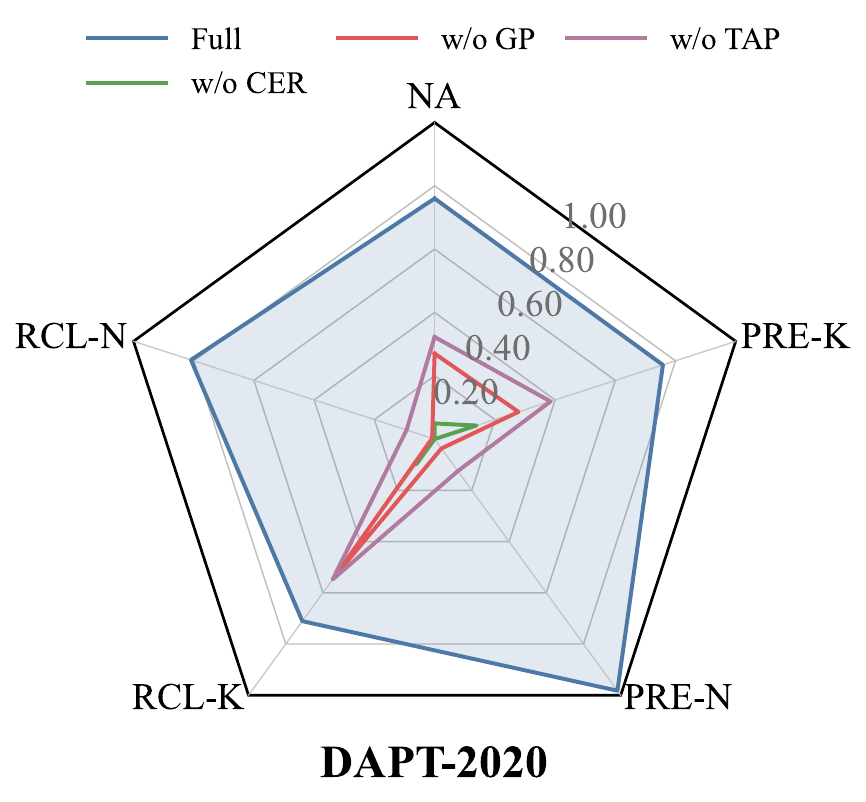}}
  \subfloat[]{\includegraphics[width=0.3\linewidth]{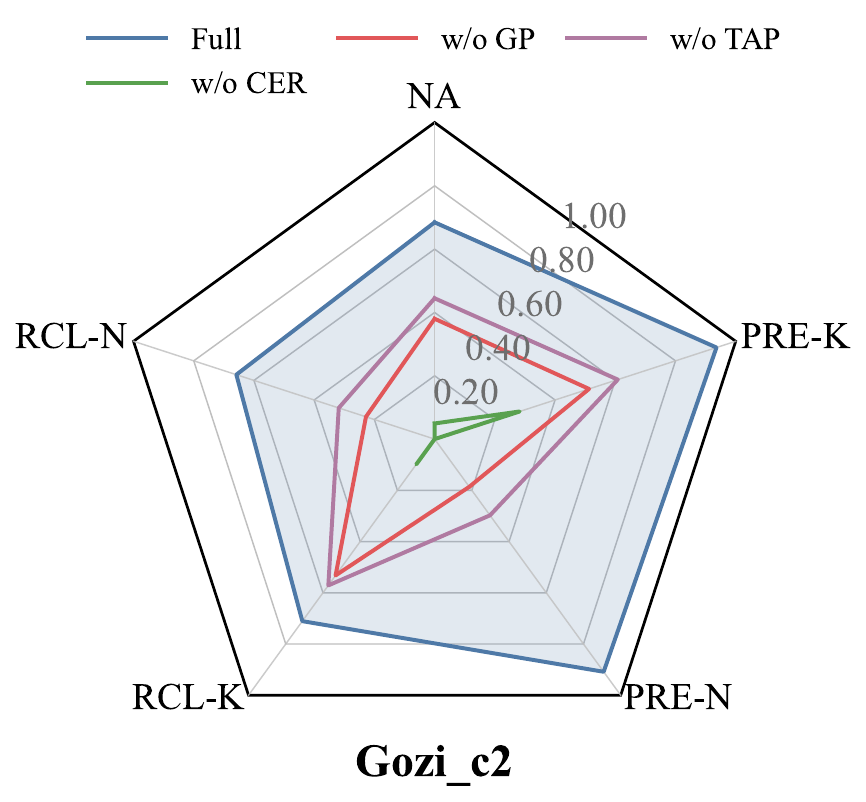}}\hfill
  \caption{Ablation results of novel malicious traffic discovery on different datasets. (a) Android Mischief dataset as novel malicious traffic, (b) DAPT-2020 dataset as novel malicious traffic, (c) Malicious\_c2 dataset as novel malicious traffic.}\label{fig:ablation_novel}
\end{figure}

We further evaluate the influence of different components on discovering novel malicious traffic.  
Figure~\ref{fig:ablation_novel} reports the results across six scenarios, each containing both known and unseen attack types.  
When the retrieval mechanism is removed, the LLM can only analyze the input flow itself without contextual evidence, leading to severe performance degradation across all metrics.  
This observation indicates that directly employing an LLM without retrieval support is insufficient for practical network security analysis, as the model lacks grounding in the underlying distribution of traffic data.  
In contrast, the guidance prompt and Traffic-Aware Adaptive Pruning modules substantially enhance discovery precision by steering the model toward traffic-relevant reasoning and filtering out spurious or weakly correlated retrieval results.  
Both modules jointly enable MalRAG to balance the identification of known malicious flows and the discovery of novel threats, effectively improving the normalized accuracy in all evaluation scenarios.

\subsection{Deep Dive Analysis}
To further understand the internal behavior and design sensitivity of MalRAG, 
we conduct a series of deep-dive experiments focusing on two aspects: 
(1) the influence of different backbone LLMs on overall performance, 
and (2) the effect of the retrieval upper bound $k$ on the accuracy and stability of traffic identification. 
These analyses provide deeper insights into how MalRAG balances model generality, retrieval quality, and computational efficiency, 
highlighting the framework’s robustness across model scales and its resilience to parameter variation.

\begin{figure}[htbp]
    \centering
    \includegraphics[width=\linewidth]{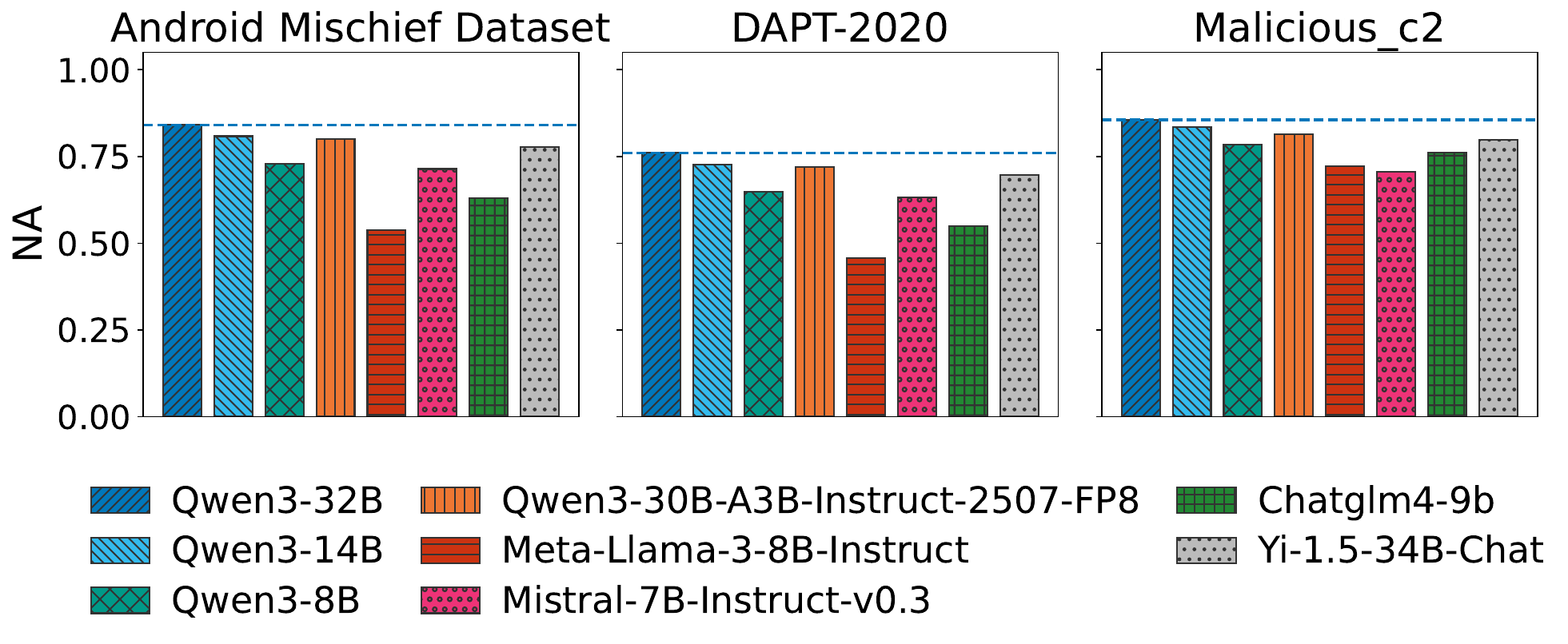}
    \caption{Impact of LLM backbone on MalRAG’s performance for novel malicious traffic discovery.}
    \label{fig:deepdive_backbone}
\end{figure}

\subsubsection{Effect of Backbone LLMs}
To examine backbone dependence, we replace the default Qwen3-32B with several alternative LLMs of varying sizes and architectures. As summarized in Table~\ref{tab:deepdive_backbone} and Figure~\ref{fig:deepdive_backbone}, MalRAG does not rely on a specific backbone: all models yield reasonable performance, though larger LLMs generally perform better. The main difference appears in recall for known malicious identification, where bigger models achieve higher recall, likely due to a stronger ability to interpret retrieved evidence. In contrast, performance gaps in novel discovery are smaller, as Traffic-Aware Adaptive Pruning removes much of the irrelevant evidence, reducing sensitivity to backbone choice and stabilizing novel detection across models.
\begin{figure}[htbp]
  \centering
  \subfloat[CTU-13]{\includegraphics[width=0.3\linewidth]{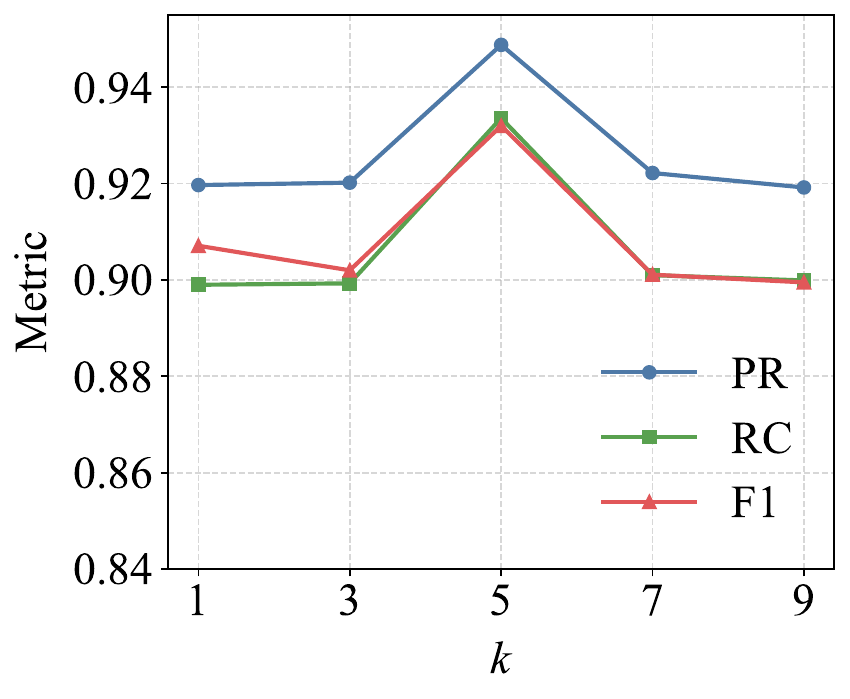}}\hfill
  \subfloat[USTC-2016]{\includegraphics[width=0.3\linewidth]{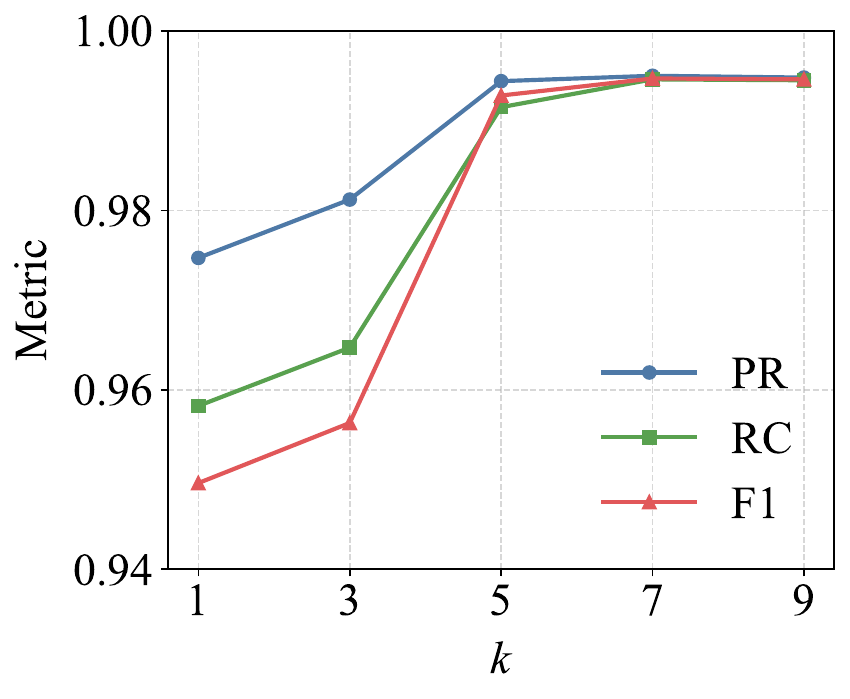}}\hfill
  \subfloat[DAPT-2020]{\includegraphics[width=0.3\linewidth]{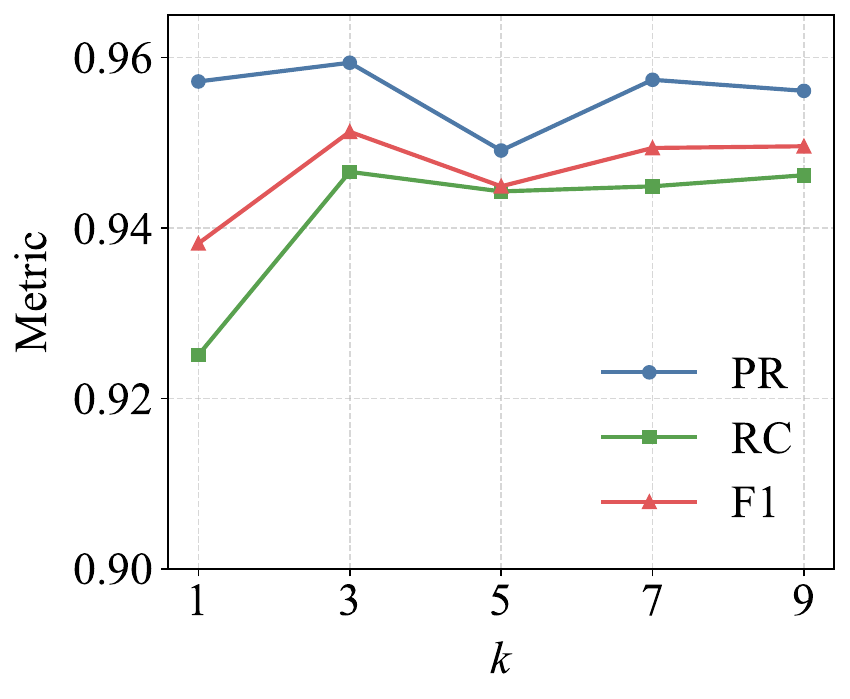}}
  \caption{Impact of retrieval upper bound $k$ on MalRAG’s performance for known malicious traffic identification.}
  \label{fig:deepdive_k}
\end{figure}
\subsubsection{Effect of Retrieval Upper Bound $k$}  
We further examine the impact of the retrieval upper bound $k$ on MalRAG’s performance.
For each query flow, the system retrieves up to $k$ nearest samples per view to construct the evidential context. Figure~\ref{fig:deepdive_k} shows results on three datasets with $k$ from 1 to 9. Performance improves steadily as $k$ increases from 1 to 5, indicating that a moderate number of highly similar samples provides richer and more consistent evidence. Beyond $k=5$, gains become marginal or slightly decline, as larger sets introduce weaker or redundant evidence that can dilute decision consistency and increase inference cost. We therefore set $k=5$ by default, balancing identification accuracy, evidential diversity, and computational efficiency.

\section{Discussion}
While our core formulation assumes a prior corpus of raw malicious traffic from which multiple complementary feature views can be extracted, real-world deployments often cannot retain such complete data. In many environments, raw packet captures are discarded and only flow-level or single-view features are stored due to privacy or storage constraints. In these settings, MalRAG can still operate by indexing the available flow statistics or other stored views and adapting the retrieval metrics, with retrieval and reasoning naturally confined to a reduced set of feature views. As an initial probe, we extracted CICFlowMeter~\cite{cicflowmeter-github} flow statistics on CTU-13 and evaluated fine-grained known MTI, obtaining an F1-score of 0.87. This suggests that performance is partly bounded by the expressiveness of statistical features, which may blur fine-grained behavioral differences, and that similarity design over such features (e.g., metric learning, feature standardization and weighting, or hybrid distances) warrants further tuning. We view this feature-only setting as a complementary deployment path for MalRAG in data-constrained environments and a promising direction for refining the similarity module.

\section{Conclusion}
In this paper, we proposed MalRAG, a retrieval-augmented LLM framework for open-set MTI. MalRAG integrates four components: a multi-view traffic database that stores comprehensive traffic knowledge; Coverage-Enhanced Retrieval to broaden relevant evidence; Traffic-Aware Adaptive Pruning to filter misleading matches; and guidance prompts that clarify the task, expose retrieved evidence, and steer the LLM toward stable decisions. Without any fine-tuning, MalRAG leverages LLM reasoning over structured traffic evidence. Extensive experiments and ablations across heterogeneous datasets show its superior performance, highlighting retrieval-augmented, training-free reasoning as a promising paradigm for intelligent network defense.




\bibliographystyle{IEEEtran}
\bibliography{reference}
 
%

\end{document}